\documentclass[graybox]{svmult}

\usepackage{mathptmx}       % selects Times Roman as basic font
\usepackage{helvet}         % selects Helvetica as sans-serif font
\usepackage{courier}        % selects Courier as typewriter font
\usepackage{type1cm}        % activate if the above 3 fonts are
\usepackage{makeidx}         % allows index generation
\usepackage{graphicx}        % standard LaTeX graphics tool
\usepackage{multicol}        % used for the two-column index
\usepackage[bottom]{footmisc}% places footnotes at page bottom

\makeindex          

\usepackage[english]{babel}
\usepackage[utf8x]{inputenc}
\usepackage[T1]{fontenc}
\usepackage{tikz}
\usetikzlibrary{fit}

\usepackage[a4paper,top=3cm,bottom=2cm,left=3cm,right=3cm,marginparwidth=1.75cm]{geometry}

\usepackage{amsmath}
\usepackage{amssymb}
\usepackage{amstext}
\usepackage{graphicx}
\usepackage{bbm}
\usepackage{natbib}
\usepackage[colorinlistoftodos]{todonotes}
\usepackage[colorlinks=true, allcolors=blue]{hyperref}
\usepackage{algorithm}
\usepackage{algorithmic}
\usepackage{hyperref}

\usepackage{booktabs}
\usepackage{array}
\newcolumntype{P}[1]{>{\centering\arraybackslash}p{#1}}

\newtheorem{hypothesis}{Hypothesis}
\newtheorem{algo}{Algorithm}

\newcommand{\zbm}{\mathbb{Z}}
\newcommand{\dbm}{\mathbb{D}}
\newcommand{\Var}{\text{Var}}

\usepackage{listings}
\usepackage{color}

\definecolor{dkgreen}{rgb}{0,0.6,0}

\begin{document}
\title*{Randomized experiments to detect and estimate\\ social influence in networks}
\titlerunning{Randomized experiments}
% your contribution title if the original one is too long
\author{Sean J. Taylor and Dean Eckles}
% your contribution title if the original one is too long
\institute{Sean J. Taylor \at Facebook \\ \email{sjt@fb.com}
\and Dean Eckles \at Sloan School of Management and Institute for Data, Systems \& Society, Massachusetts Institute of Technology  \\ \email{eckles@mit.edu}}

\maketitle
\abstract{Estimation of social influence in networks can be substantially biased in observational studies due to homophily and network correlation in exposure to exogenous events.  Randomized experiments, in which the researcher intervenes in the social system and uses randomization to determine how to do so, provide a methodology for credibly estimating of causal effects of social behaviors. In addition to addressing questions central to the social sciences, these estimates can form the basis for effective marketing and public policy.
\newline \indent In this review, we discuss the design space of experiments to measure social influence through combinations of interventions and randomizations.  We define an experiment as combination of (1) a target population of individuals connected by an observed interaction network, (2) a set of treatments whereby the researcher will intervene in the social system, (3) a randomization strategy which maps individuals or edges to treatments, and (4) a measurement of an outcome of interest after treatment has been assigned.  We review experiments that demonstrate potential experimental designs and we evaluate their advantages and tradeoffs for answering different types of causal questions about social influence. We show how randomization also provides a basis for statistical inference when analyzing these experiments. \newline \newline 
\textbf{Keywords} Field experiments, causal inference, Fisherian randomization inference, social interactions, spillovers, social networks}

\section{Introduction}

There is a long tradition in the social sciences of examining how individual level behaviors diffuse and aggregate, including influential work by \cite{schelling1969models,schelling1971dynamic,schelling1973hockey} and \citet{granovetter1978threshold}, among many others \citep{ryan1943diffusion, bass1969ms, valente1996social,morris2000contagion}.  Stylized models from this tradition have been used to explain some of the most important human phenomena, from which innovations are likely to gain widespread usage to who people vote for in elections. The fundamental building blocks of diffusion models are assumptions about how people change their behaviors in response to the behaviors of people they observe or interact with.  These assumptions can vary in their disciplinary origins and sophistication --- from epidemiological models to game-theoretic models with multiple equilibria.

Randomized experiments provide a useful tool for testing theories.  The increasing digitization and connectedness of human behaviors has made digital field experiments cheaper and easier to apply to social behaviors via contemporary communication technologies.  This methodological paradigm shift has created opportunities for researchers hoping to understand the underpinnings of large-scale social behaviors in order to improve theory, make predictions, and compare hypothetical policies.

In this review, we hope to make randomized experimentation more accessible to researchers seeking to contribute to our understanding of mechanisms for social influence and diffusion in social systems.  We first discuss how randomized experiments can rule out potential confounding factors (Section~\ref{sec:different}).  Because experiments require that the researcher intervenes in the social system, we devote Section~\ref{sec:ethics} to discussing the ethical consideration associated with employing digital field experiments.

Section~\ref{sec:components} outlines the four components of a randomized experiment to detect or estimate social influence. This facilitates discussing the many design choices experimenters have, including defining the relevant network, what treatments can be employed, and how those treatments may be randomly assigned to subjects.   In Section~\ref{sec:analyzing}, we turn to the analysis of experiments in networks, where we focus on Fisherian randomization inference.  Section~\ref{sec:interpreting} discusses how the analysis of experiments can be extended in various ways in order to increase the usefulness of the results.

This review complements more general references on design and analysis of randomized experiments \citep{gerber2012field,athey2017econometrics,imbens2015causal}. Design and analysis with disjoint groups, has received substantial attention in economics and epidemiology \citep[e.g.,][]{saul2017causal,halloran2016dependent,baird2016optimal,vanderweele2012components}.
On the other hand, there are few other reviews of design and analysis of experiments in networks. Compared with extant reviews \citep{aral_networked_2016,walker_design_2014}, we aim to integrate all the methods reviewed into a single causal model and discuss some design choices and analysis methods in detail.

What exactly counts as social influence? Different fields distinguish among various processes by which people affect each other. For example, economists distinguish between peer effects caused by constraint, preference, and expectation interactions \citep{manski2000economic}, while other fields may make different distinctions. Thus, for some prior work, ``social influence'' denotes something more specific. However, given our methodological focus here, we choose to remain agnostic about the mechanisms and define \emph{social influence} to include all processes by which an individual's behaviors affect another's, either directly or indirectly. Thus, we could have instead referred to ``peer effects'', ``diffusion'', or ``social contagion''. Further theory-specific distinctions may motivate additional design and analysis choices.

\subsection{What makes randomized experiments different?}
\label{sec:different}

We privilege information gained through randomized experiments because they create a different kind of knowledge than observational studies. In particular, because we know exactly how units are assigned to treatments, a properly implemented experiment rules out all alternative explanations for an observed correlation besides the causal one, and allows for both unbiased estimation of the effect of our intervention and statistical inference that is exact in finite samples \citep{fisher1935design}.

Any observational analysis intended to answer questions about social influence must confront several potential biases that make it difficult to trust its conclusions.  First, social networks are known to exhibit strong homophily \citep{mcpherson2001birds,marmaros2006friendships,currarini2010identifying,kossinets2006empirical}, creating network correlation of attributes, opinions, and behaviors through people's preferences for whom they spend time with.  For instance, people with similar political beliefs may be more likely to form friendships \citep{bakshy2015exposure,halberstam2016homophily,huber2017political}, and therefore homophily may readily explain cases of apparent political persuasion.  Second, people who are connected in social networks are subject to similar exogenous shocks to their behavior, as when neighbors are exposed to similar marketing messages on billboards.

A substantial program of research has been devoted to proving that what economists call ``identification problems'' in social influence are likely to be insurmountable without randomized experiments \citep{manski1993identification,shalizi2011homophily}.  The intuitive reason is that without intervening in the social system, there are usually reasonable alternative explanations for correlations that do not involve a social influence effect.

Despite their clear advantages, the use of randomized experiments is not a panacea for social scientists.  Experiments are usually more costly to design and implement than observational studies because the researcher must alter people's behaviors or interactions in social system in some way.  Interventions require substantial upfront costs for planning and implementation, including: recruitment of subjects, cost of the interventions themselves (financial or logistical), evaluation and exposure of risks of harm to subjects \citep{jackman2016evolving}.  Because field experiments require researchers to impact the social systems they study at potentially very large scale, they can be associated with larger or different ethical challenges from other methods, which we summarize in Section~\ref{sec:ethics}.

Experiments can also be problematic because, although they reduce concerns about bias, the variance of estimation becomes a first-order concern and the possibility of type II errors (commonly known as issues with experimental power) dominate due to the cost of sample size or the impossibility of the researcher creating large effects \citep{bakshy2013uncertainty}.

On a more positive note, we will see in Section~\ref{sec:analyzing} that some well-designed experiments can require more straightforward analysis than observational studies.  In addition, there are no internal validity concerns with experiments -- they provide unbiased estimates for the social influence effects they were designed to measure. The two main constraints of an experimental methodology are which estimates are possible and the precision of those estimates.

The randomization the researcher employs and structure of the network together determine what causal quantities of interest can be credibly estimated \citep{toulis_estimation_2013}.  These estimands address counterfactual questions about which individual-level behaviors would obtain under alternative interventions. In one simple case, we may be able answer the question of how much an individual's probability of a behavior is increased by having exactly one peer (rather than no peers) who engages in that behavior.  A more complex causal estimand might be the distribution of that behavior in the total population after a series of targeted (e.g., marketing) interventions or a policy change (e.g., by a government).  As we will see, no single experiment can answer all possible causal questions and the experiment should be designed with some estimands in mind.

The second constraint from using experiments is the precision of the effect estimates.  Experimental data is often more costly to collect than observational data because treatments are not free and because observational data is abundant.  In general, the precision of the estimates of social influence are limited by the direct effects of the intervention (weaker ones provide less experimental power) and the available sample size \citep{gerber2012field}.

\subsection{Ethical considerations for digital field experiments}
\label{sec:ethics}

\begin{table}[t]
  \caption{Ethical principles for human subjects research.}
  \begin{tabular}{p{4cm} p{9cm}}
  \toprule
  Principle & Description \\
  \midrule
  Respect for Persons & Treating people as autonomous and acting in accordance with their wishes. \\[3pt]
  Beneficence & Recognizing the potential risks and benefits of research and striking a balance between them. \\[3pt]
  Justice & Ensuring that the risks and benefits of research are fairly distributed. \\[3pt]
  Respect for Law \\and Public Interest & Recognizing the risks and benefits for all relevant stakeholders, not just research subjects.\\[3pt]
  \bottomrule
  \end{tabular}
  \label{tab:principles}
\end{table}

The reduced cost and increased feasibility of digital field experiments (DFEs) has led to increased experimentation over the past decade.  While DFEs may help researchers answer many important questions about social influence, they can present more ethical challenges than observational research and even pre-digital lab and field experiments.  To ground the discussion, we will refer to the four ethical principles proposed in the Belmont Report \citep{national1978belmont} and the subsequent Menlo Report \citep{dittrich2012menlo} which are meant to provide guidance on human subjects research.  Those principles -- Respect for Persons, Beneficence, Justice, and Respect for Law and Public Interest -- are briefly summarized principles in Table~\ref{tab:principles}.

For an in-depth, thorough treatment of ethics in research in the digital age, we refer the reader to Chapter 6 of \cite{salganik2017bit} and for a recent discussion of institutional review processes to mitigate risk please see \cite{jackman2016evolving}.  Rather than review those materials exhaustively, we use this subsection to discuss five ethical considerations that we consider to be particularly salient for digital field experiments.  

First, DFEs are implemented in software and therefore have very low variable costs with respect to the size of the treated population. It is no longer unusual for experiments to deploy treatments to millions of people \citep{bond_experiment_2012}, amplifying their potential harm compared to more modest sample sizes.  Additionally, treatments with network effects can, by research intention or not, cause detrimental effects for people who were not in the original treated population.  Researchers acting in accordance with the ethical principle of Beneficence may have a more difficult time evaluating the potential risks of DFEs in networks because their potential effects on social systems are not obvious, intuitive, or even measured.

Second, it may be difficult to identify whether subjects in DFEs are members of a vulnerable or protected population.  When designing a DFE, researchers might find it challenging to estimate risks of harm because there is uncertainty about how many subjects could be adversely affected.  Researchers might also be unable to reason about whether the benefits and risks of the research are distributed equitably across the population, in accordance with the principle of Justice.   On many online or mobile platforms, researchers may not know if users are a reasonable age for consent or are particularly vulnerable to risk from the planned experiment.

Third, DFEs typically use automated, large-scale collection of potentially sensitive and/or identifying information, e.g. location information or exchange of personal communication.  Indeed, these data can be integral to the ability of the experiment to answer the research question of interest.  For instance, a log of email communications can be used to infer a social network \citep{kossinets2006empirical}, which is a key component for social influence studies.  Persistent records of sensitive or identifying information can potentially be used for unintended purposes, causing harm to experimental subjects \citep{ohm2010broken,narayanan2010myths}.

Fourth, because of their large scale and integration with existing technologies, DFEs often pose unique challenges for receiving informed consent, which is sometimes an implication of the ethical principle of Respect for Persons.  Informing subjects of the experiment and receiving their consent can be disruptive to their normal experiences using various platforms and products (particularly if experiments are frequent, as is becoming more common).  Furthermore, requiring informed consent can limit or bias the experimental population or prime the subjects, undermining or altering the treatment effects.  Although informed consent is important component of Respect for Persons, deception may be permissible if the experiment complies with all other ethical principles and the deception does not strongly violate the norms of that setting \citep{riach2004deceptive}. Some experiments with potentially important benefits \emph{require} deception in order ensure the research question can be suitably answered.  For instance, in the employment discrimination field experiments \cite{riach2004deceptive} discuss, one could not credibly measure discrimination after informing employers of the nature of the research.

Fifth, it may be difficult for researchers to comply with all laws, contracts, terms of service, or social norms because DFEs may involve partnerships with companies, span countries or other legal boundaries, or include subjects from many cultures.  Inconsistent, overlapping, and sometimes unclear rules and norms lead to challenges for researchers hoping to understand all potential stakeholders and their associated goals and risks.

These five considerations are not meant to be exhaustive -- there are certainly other ways in which DFEs can present new ethical challenges for researchers.  But we hope that this subsection has made clear that while experimental research has become easier to conduct on some dimensions, it has become more fraught on others -- in particular in evaluating and mitigating the risk of harm to subjects.

\subsubsection{Recommendations for ethical research}

Taking into account the challenges identified, more research is needed to address the ethical implications in DFEs and to develop mitigating and creative strategies. In the meantime, researchers should do the utmost to:

\begin{itemize}
\item Ensure that the research is ethical and beneficial for subjects; that it does not expose them to risk or harm (this may require escalation and further deliberation with other teams within the company, along with the assessment of alternative research methods that could be used). 
\item Carefully assess if the collection and processing of sensitive data is essential for the research being conducted.
\item Determine if an experiment is strictly necessary for the objectives of the research (or if the same results can be obtained through less risky research, e.g. a smaller sample size).
\item Whenever possible, ensure that such collection and processing is done with prior informed consent given by the data subjects.
\item Only keep that data for the minimum necessary period of time and
 ensure the proper de-identification of that data according to most effective and updated industry standards.
\end{itemize}

\section{Components of a randomized experiment}
\label{sec:components}

The randomized experiment methodology has four main components:

\begin{enumerate}
\item A \textbf{target population} of units (i.e. individuals, subjects, vertices, nodes) who are connected by some \textbf{interaction network}.  (Section~\ref{sec:target})
\item A \textbf{treatment} which can plausibly affect behaviors or interactions. (Section~\ref{sec:treatments})
\item A \textbf{randomization strategy} mapping units to probabilities of treatments. (Section~\ref{sec:randomization})
\item An \textbf{outcome} behavior or attitude of interest and measurement strategy for capturing it. (Section~\ref{sec:outcomes})
\end{enumerate}

To summarize the relationship of these components, the researcher applies a treatment (2) to a target population (1) using a randomization strategy (3) and then measures the outcome behavior (4).

The following four sections describe these four components, characterize the space of possibilities for each one, and provide examples from existing research.  We introduce notation along the way that we will use in Sections~\ref{sec:analyzing} and \ref{sec:interpreting}.  For convenience that notation is summarized in Table~\ref{tab:terms}. Lowercase letters designate particular fixed values of interest.

\subsection{Target population and interaction network}
\label{sec:target}

\begin{table}[t]
  \caption{Definitions of terminology and notation used in this review.  See Figure~\ref{fig:diagram} for a graphical depiction of the relationship between these quantities.}
  \begin{tabular}{p{2cm} p{9cm}}
  \toprule
  Term & Definition \\
  \midrule
  $X_i$ & a vector of pre-treatment covariates about subject $i$ \\[3pt]
  $U_i$ & a vector of unobserved covariates about subject $i$ \\[3pt]
  $D_j$ & a behavior of the peer $j$ that could affect the subject $i$ \\[3pt]
  $A_{ij}$ & edge between $i$ and $j$ in the interaction network that mediates social influence \\[3pt]
\hline
  $Z_j$ & researcher-determined treatment status for peer $j$.  \\[3pt]
  $W_{ij}$ & researcher-determined treatment status for relationship $ij$ \\[3pt]
\hline
  $Y_i$ & the outcome of interest for subject $i$, measured post-treatment \\[3pt]
\hline
  subject & a focal individual whose outcome variable $Y_i$ is studied \\ [3pt]
  peer & the person whose behavior $D_j$ could influence $Y_i$ \\[3pt]
  \bottomrule
  \end{tabular}
  \label{tab:terms}
\end{table}

\begin{figure}[t]
\label{fig:diagram}
\centering
\begin{tikzpicture}[
round/.style={circle, draw=black!60, very thick, minimum size=10mm},
square/.style={rectangle, draw=black!60, very thick, minimum size=10mm},
]
%Nodes
\node[square]      (outcome) {$Y_i$};
\node[square]      (covariates) [above=of outcome] {$X_i$};
\node[round]       (unknown) [above=of covariates] {$U_i$};
\node[square]      (friendship) [above left=of outcome] {$A_{ij}$};
\node[square]      (friendbehavior) [left=of outcome] {$D_j$};
\node[square]      (friendtreatment) [left=of friendbehavior] {$Z_j$};
\node[square]      (friendcovariates) [above left=of friendbehavior] {$X_j$};
\node[round]       (unknownfriend) [above=of friendcovariates] {$U_j$};
\node[square]      (friendshiptreatment) [above=of friendship] {$W_{ij}$};

%Lines
\draw[->] (unknown.south east) to[out=-45,in=45] (outcome.east);
\draw[->] (unknownfriend.south west) to[out=-135,in=-135, distance=4cm] (friendbehavior.south west);
\draw[dotted] (unknown.north west) to[out=135,in=45, distance=2cm] (unknownfriend.north east);
\draw[->] (covariates.west) -> (friendship.east);
\draw[->] (friendcovariates.east) -> (friendship.west);

\draw[->] (unknownfriend.south east) -> (friendship.north west);
\draw[->] (unknown.south west) -> (friendship.north east);

\draw[->] (covariates.south) -> (outcome.north);
\draw[->] (friendbehavior.east) -> (outcome.west);
\draw[->] (friendship.south east) -> (outcome.north west);
\draw[->] (friendshiptreatment.south) -> (friendship.north);
\draw[->] (friendcovariates.south east) -> (friendbehavior.north west);
\draw[->] (friendtreatment.east) -> (friendbehavior.west);

% Boxes
\node[dotted, draw=red, fit=(friendshiptreatment) (friendship), inner sep=2mm](b2) {};
\node[dotted, draw=red, fit=(friendtreatment) (friendbehavior), inner sep=2mm](b3) {};

\end{tikzpicture}
\caption{Causal diagram for the random variables in our example.  Squares are observed variables and circles are unobserved.  Here $i$ is the ego or focal subject for whom we will measure outcome $Y_i$.  We believe that her friend $j$ can affect $Y_i$ through her behavior $D_j$, which is affected by our treatment $Z_j$.  The strength of their friendship, $A_{ij}$ can moderate this effect and is exogenously affected by treatment $W_{ij}$. $U_i$ and $U_j$ are unobserved confounders that cause $i$ and $j$ to become friends and may also affect $D_j$ and $Y_i$.  By conditioning on $A_{ij}$, the backdoor path indicated by the dashed line is activated and provides an alternative explanation for any association we observe between $D_j$ and $Y_i$.}
\end{figure}
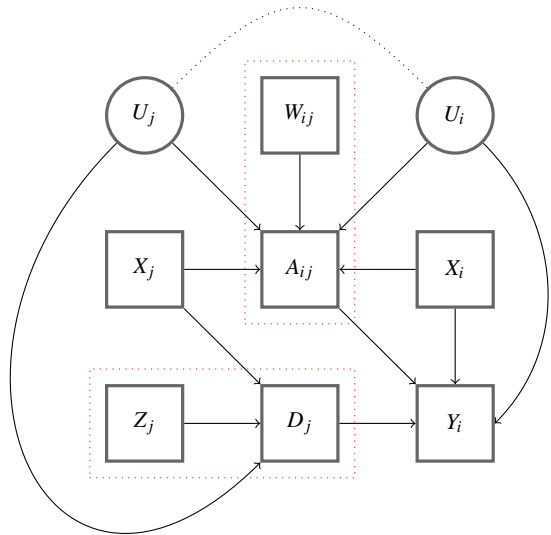

The target population is the set of people whose interactions and behaviors the researcher seeks to study.  If we were studying whether peers affect which movies we choose to watch, the population of interest might be movie-goers.  Before and during the experiment the target population generates some data, which we list here:

\begin{itemize}
\item We observe $N$ individuals from some \textbf{target population}, indexed by $i$. This might be a sample or it may be the entire finite population.
\item We observe \textbf{pre-treatment covariates} for the individuals: $X_i$.  Commonly researchers collect demographic information such as gender, physical location, or age.  Often it is also useful to measure pre-experimental behaviors that are similar to the outcome of interest.
\item We observe an \textbf{interaction network} between people in the population $A_{ij}$ where $i, j \in [1, \dots, N]$; alternatively, this is a network $G = (V, E)$.  This interaction network determines an exposure model --- which individuals we expect to potentially influence each other and with what intensity.
\item We observe when the population engages in some behavior of interest $D_i$.
\item We observe some \textbf{outcome variable} associated with each individual, $Y_i$.  For instance the researcher might survey them to ask often they smoke.  We will discuss outcome measurement in more depth in Section~\ref{sec:outcomes}.
\end{itemize}

Substantively, we care about the effect of the behavior $D_j$ on the outcome $Y_i$ in the population.  The special case where $D_i = Y_i$ can be termed \emph{in-kind peer effects} and is frequently studied, but it is easy to envision cases where the peer behavior of interest is different from the outcome (e.g. my friend's studying habits, measured as $D_j$, affect my probability of applying to college, $Y_i$).

Selecting the target population often involves tradeoffs between external validity and the ability to collect data about behaviors, outcomes of interest, and relevant social interactions --- and intervene. Researchers have used the following three strategies to solve this \emph{recruitment} problem.

First, researchers have continued to recruit convenience samples. As with classic lab experiments in social influence \citep[e.g.][]{asch1955opinions}, these are often students from universities and colleges for studies. These samples can facilitate either construction of artificial networks or measuring the subjects' networks (with, e.g., surveys, asking them to log into Facebook, measuring co-location). The later strategy can be used to conduct ``lab experiments in the field'' as existing networks are combining with artificial choices and treatments \citep[e.g.,][]{leider2009directed}. The former strategy has been increasingly been used in combination with online labor markets (such as Amazon's Mechanical Turk), which has created an important new source of experimental subjects \citep{mason2012conducting}. These individuals can be assigned to positions in networks by researchers or through economic games played by the subjects themselves \citep{suri2011cooperation,mason2012collaborative,rand2011dynamic,rand2014static}; of course, this may limit external validity.

Second, the last decade has led to a dramatic increase in experiments that are conducted on online social networks or in collaboration with the companies that run online communication services.  \cite{aral2011creating,aral2012identifying} constructed a Facebook application in order to gather social network information, introduce a treatment (presence of viral features), and measure the outcome of interest (adoption of the application).  Other researchers have worked directly with Internet firms to conduct experiments.  \cite{bond_experiment_2012} and \cite{taylor2013selection} conducted experiments by implementing them in partnership with Facebook (see Bond et al., this volume), while \cite{muchnik2013social} partnered with a social news website to introduce an experimental change.

Third, researchers in education, development, labor economics, and ecology have conducted ambitious field experiments in samples of schools or classrooms \citep{carrell_does_2009, paluck_changing_2016}, villages \citep{kim2015social,cai_social_2015}, and herds of animals \citep{aplin_experimentally_2015, firth_pathways_2016} for which networks can be measured.

\subsubsection{Measuring or constructing the interaction network}

We use the term ``interaction network'', which is vague, because what is usually denoted by ``social network'' will often not be the causal network of interest.\footnote{Another related term is ``exposure model'' -- a model that determines which subjects are exposed to which other subjects.}  In most settings there is some specific type of interaction we hypothesize to transmit the behavior we care about.  The most intuitive definition is that interaction is ``$i$ considers $j$ to be her friend,'' but, even when this can be operationalized, further consideration of a particular research question may lead to other reasonable choices:

\begin{itemize}
\item $i$ saw a story $j$ posted on Facebook \citep{bakshy_role_2011}
\item $i$ is made aware that her friend $j$ likes a product \citep{bakshy_social_2012}
\item $i$ lives with $j$ in a dormitory for a year \citep{sacerdote2001peer}
\item $i$ lives in the same household as $j$ \citep{nickerson2008voting}
\item $i$ is in the same training class as $j$  \citep{carrell_does_2009}
\end{itemize}

The researcher hopes that the chosen network captures salient interactions for the influence process she expects.  This definition can vary depending on the outcome behavior of interest. In the case of the \cite{sacerdote2001peer}, who study educational outcomes, the interaction network is prolonged co-habitation, while in the case of \cite{bakshy_social_2012}, who study clicks on ads, it is merely that a Facebook friend's name can appear next to an advertisement.  A more prolonged, socially important interaction network can plausibly cause larger changes in subject behavior. In the former case, the the researchers can study changes in more important and ingrained behaviors like studying habits, while in the latter the researchers must study more proximate outcomes (clicks on ads).

There are many different possibilities for measuring, eliciting, or directly constructing interaction networks.  If the research setting is an articulated social network, (e.g. an online social network such as Facebook, Instagram, Twitter, or Pinterest) the researcher may use that network's definition (followers, friends, subscriptions).  This approach is convenient but often not the precise interaction network of interest.  Most people have online ``friends'' with whom they never interact in person, as well as ``real life'' friends who they have not articulated ties with online.  Facebook, Instagram, and Twitter use algorithmic ranking to determine which content users see, meaning that a friend or follow relationship on those platforms may not necessarily imply content visibility.  If the plausible mechanism of influence is offline, then using an online network might bias estimates of causal effects.  A misspecification of the interaction network can, even with randomization, bias measurement of social effects.

In digital settings, interaction networks may be constructed incrementally as people's interactions in the social system are logged (i.e. $A_{ij} = 1$ if $i$ chatted with person $j$ during the study). For instance in \cite{bakshy_social_2012}, the interaction network is determined by Facebook users seeing advertisements during their browsing sessions.  The salient interaction network is easily captured by logging which users see which ads.  In addition, logging the interactions which have the potential for transmitting  behaviors can improve precision by omitting interactions with no potential to transmit influence. \cite{bakshy_social_2012} could have used other definitions of the interaction network (e.g. Facebook friendship), but these would have yielded biased and/or higher variance estimates of effects.

Like observational research [e.g., the US National Longitudinal
Study of Adolescent Health (AddHealth) study \citep{resnick1997protecting}], much measurement of social networks for randomized experiments has involved asking subjects who their friends, kin, etc., are. The specific questions can be selected to elicit the possibly domain-specific network of interactions. For example, \citet{cai_social_2015} asked heads of rural households to household heads to list five friends that they most frequently discuss farming and finance with, anticipating that this would be a relevant network for social influence in adoption of weather insurance and spillovers from their intervention. Such questions require being able to uniquely identify the named peers, which may be challenging in the presence of common names and/or limited literacy. \citet{kim2015social} thus used a complete photographic census of the villages in which they planning to intervene. When the goal is to measure an objective fact about behavioral interactions, incentives for subjects to truthfully report their friends and tie-strength to researchers could be helpful. For example, \citet{leider2009directed} use a game in which individuals report how much time they spend with peers and paying them more money if this report matches the peer's report.

Researchers can infer interaction networks from communication meta-data, especially when it covers enough time to precisely measure interaction rates and the communication medium (e.g. email) is likely to be the medium through which influence is transmitted. Influential observational research has measured networks by counting exchanges of emails \citep{kossinets2006empirical} or instant messages \citep{aral2009distinguishing}.
Beyond allowing for constructing a binary network, directed behaviors between individuals predict self-reported tie strength \citep{jones2013inferring}. In a a randomized experiment, these measures can then be used to estimate how spillovers \citep{bond_experiment_2012} or social influence \citep{bakshy_social_2012,bakshy_role_2011,aral2014tie,bond2016social} vary by tie strength.
Choices by researchers in inferring networks from communications data can be non-trivial and have a substantive impact on results \citep{de2010inferring}.

Finally, studies can be designed to \emph{directly construct} the interaction network for the subjects, a strategy which is enabled by running digital experiments even if they happen to be conducted synchronously in behavioral research labs \citep{kearns2006experimental,mason2012collaborative,rand2011dynamic,rand2014static}. For example, \cite{suri2011cooperation} randomly vary the networks on which Amazon Mechanical Turkers play a public goods game.  Since creating the interaction network requires the researcher to intervene in the social system, we will discuss this strategy in more depth in Section~\ref{sec:treatments}.

\subsubsection{Extensions to this framework}

Thus far we have described a randomized experiment with a single time period of post-treatment observation and a single outcome of interest.  The DAG in Figure~\ref{fig:diagram} does not allow for the subject's behavior to affect the peer's behavior, which in turn affects the subject's behavior. There are two simple extensions which may be useful and more realistic.  First, we might study the outcome at different points in time (e.g. instead of $D_i$ and $Y_i$ we might observe $D_i(t)$ and $Y_i(t)$ where $t$ denotes either discrete or continuous time.  Time-dependent behavior is a challenging empirical setting because the researcher will often need to model how the interaction network varies across time, as well as how the individual behavior evolves over time \citep{rock2016identification}.  %Often it is convenient to segment the experiment to two periods, pre- and post-treatment, and use the additional time-variation in treatment status to help identify causal effects.

The second extension is from a single peer behavior and outcome of interest to multiple behaviors and outcomes.  We might observe a set of people make decisions about a collection of products, ads, content items, or behaviors, meaning we would measure $D_{ik}$ and $Y_{ik}$, where $k$ indexes the items.  Multiple items present an important opportunity to observe social influence processes play out repeatedly in the same population of individuals across the same interaction network. Studies which measure effects across multiple behaviors might provide a more generalizable estimate of effects or allow the researcher to understand effect heterogeneity on other dimensions. As we discuss in Section \ref{sec:analysis_other}, this may offer additional opportunities in analysis. 

\subsection{Experimental treatments} \label{sec:treatments}

Treatments are the means by which the researcher intervenes in the social system.  The space of treatments is often very limited based on cost and practical constraints, risks to subjects, and simply what changes a researcher can possibly apply in a social system.

We will consider the researcher intervening by setting variables $Z_j$ and $W_{ij}$, usually through some random assignment procedure. Note that we do not assume the researcher can directly change $D_j$ and $A_{ij}$, as these variables are chosen by individuals and can often only be affected through the researcher-controlled instruments.  The case where this is possible is the special case of perfect compliance, which is rare in field experiments.  Instead, we posit a (potentially estimable) compliance model that produces $D_j$ and $A_{ij}$ and which may also include pre-treatment variables and random noise. This section focuses on defining these treatments; we defer their random assignment to Section \ref{sec:randomization} below.

\subsubsection{Subject-level treatments}

A binary subject-level treatment is denoted by $Z_j \in \{0, 1\}$, where $Z_j = 0$ by default, and where this treatment is expected to affect behavior such that $D_j(z_j) = f_i(z_j, \epsilon_j)$, with observed $D_j = f_i(Z_j, \epsilon_j)$. The direct effects of the treatment may sometimes be of interest (e.g., effects of a message on voter turnout), but the idea here is that $Z_j$ functions as an encouragement or instrumental variable with respect to $D_j$, allowing interpretation of spillovers from treatment as social influence via $D_j$. Thus, researchers can create these treatments primarily for this purpose of detecting social influence.
For the treatment to be effective as an instrument, we must believe that $f_i$ is such that changing $Z_j$ sometimes changes $D_j$; for example, perhaps $D_j(z_j) = \mathbbm{1}\{\alpha + \beta z_j + \epsilon_j > 0\}$ with $\beta \neq 0$, which can be tested. Many interventions (e.g., providing information, advertisements) cause only small changes changes in the behavior, making detecting downstream social influence difficult.

The special case where $D_j = Z_j$ is known as perfect compliance. Noncompliance may also be only one-sided, such that if $Z_j = 1$ then $D_j = 1$. Say we are interested in social influence in adoption of a paid upgrade of an music streaming service. We could, as do \cite{bapna2015your}, purchase the upgrade for a active users at random, thus producing one-sided, rather than two-sided, noncompliance (i.e. users could still purchase the upgrade on their own if we did not).\footnote{Of course, in such cases we may wonder whether $D_j$ (i.e. having the upgrade) was really the behavior we were interested in. Perhaps so --- if most of the effects of peers' upgrades on subjects would be via a single indicator on the peers' profiles that they had upgraded.}
Two-sided noncompliance seems to be more common in the social sciences, particularly among the difficult-to-change behaviors which are often most interesting to study (e.g., health behaviors, costly product purchases).

Experiments using subject-level treatments within groups (i.e., networks consisting of disconnected cliques) to detect and estimate social influence --- sometimes called \emph{partial population experiments} \citep{moffitt2001policy} --- have been adopted in economics and political science \citep{angelucci_indirect_2009,duflo_role_2003,forastiere_identification_2015,miguel_worms:_2004,nickerson2008voting}. These designs are based on the expectation that treating a fraction of subjects can induce detectable changes in the population of individuals connected to them. A smaller number of such experiments have been conducted in networks; these too have often relied on having a network multiple connected components (e.g., villages, schools) \citep[e.g.][]{paluck_changing_2016,kim2015social,cai_social_2015,coppock_treatments_2015} with few exceptions \citep{bond_experiment_2012}.

Knowledge of the interaction network can be crucial for the success of subject-level treatments.  If there is uncertainty about \emph{which} peers may be affected by a subject's treatment, then detecting effects can become more burdensome from a statistical standpoint because omitting edges or including irrelevant ones adds additional random variation in estimation.

%Subject-level treatments can be based on some pre-treatment covariates and network structure, and this may be useful in at least three ways.  First, the researcher may design the randomization to create covariate balance so that the treatment and control groups are more similar on pre-treatment characteristics than would be expected at random, called ``blocking'' or ``pre-stratification'' c.f. \citep{moore2012multivariate}.

%Third, the treatment may be applied differently for different parts of the network.  As we discuss in Section~\ref{sec:subject_randomization}, a clustered randomization would involve partitioning the network into clusters and then applying the randomization with different probabilities to different clusters in order to induce variance in the number of treated peers \citep{ugander_graph_2013,eckles_design_2017,baird2016optimal}.  See Figure~\ref{fig:randomization_diagram} for an illustration of this idea.

\subsubsection{Interaction-network treatments}

In an interaction-network treatment, the researcher intervenes by setting $W_{ij}$, which affects the interaction network of the subjects in the experiment; that is, $A_{ij}(w_{ij}) = g_{ij}(w_{ij}, U_i, U_j, \nu_{ij})$, with observed $A_{ij} = g_i(W_{ij}, U_i, U_j, \nu_{ij})$. As above, if $A_{ij}$ is binary we may posit that $A_{ij}(w_{ij}) = \mathbbm{1}\{\gamma + \delta w_{ij} + \nu_{ij} > 0\}$ with $\delta \neq 0$. Then particular edges may exist ($\delta > 0$) or not ($\delta < 0$) because of the treatment. 

In the edge-formation case of $\delta > 0$, we have treatments such as suggesting that two people become friends or introducing them \citep{backstrom2011supervised,schultz2012methods}. Not all suggested edges will form, but we expect that some will. 
Researchers sometimes define the interaction network such that there is perfect compliance. 
There are numerous examples of randomized \emph{group formation} with ostensibly perfect compliance. \cite{hasan2017conversational} use a novel group randomization to understand how the constituents of groups affect ideation.  \cite{sacerdote2001peer} and \cite{carrell_does_2009} use random assignment of college roommates and squadrons in order to understand how these groups affect various learning and development outcomes.  Note that the degree of ``compliance'' depends on how the network is defined.  Although roommate assignment creates perfect compliance for the network of \emph{roommates}, there is still two-sided noncompliance for the network of \emph{friendships}.

Encouraging edge removal, preventing formation, or attenuating interaction ($\delta < 0$) can also be possible, if challenging in practice, and would rely on the researcher discouraging at least one type of interaction between individuals in the population.  As an extreme example, researchers studying smoking cessation could ask subjects to delete phone contacts for any friend they believe might encourage them to continue smoking.

In the context of online communication technologies, whether some binary treatment should be understood as encouraging or discouraging interaction is relative to an arbitrary and temporary status quo. For example, \citet{eckles_estimating_2016} analyze an intervention that modifies the display of $i$'s posts to $j$, varying the salience of the user interface elements for commenting on the post.

Perfect compliance, or at least one-sided noncompliance, can also occur when there is some exhaustive channel by which interaction occurs. For example, \cite{aral2012identifying} randomize along which edges notification for their Facebook application are sent, thus defining an interaction network that is a random subset of the Facebook friendship network. Similarly, \cite{bakshy_social_2012} randomize whether whether or not a friend appears as social context for an advertisement. We have elsewhere called these \emph{mechanism} experimental designs since they randomize whether particular mechanisms for social influence are active \citep{eckles_estimating_2016}.

%\citet{becker2017network} \citet{centola2010spread}
%\citet{fowler2010cooperative}

\subsection{Randomization strategy} \label{sec:randomization}

A randomization strategy $\phi$ specifies a probability distribution over treatment assignments; here $\pi_\phi(Z)$ or $\pi_\phi(W)$, where $Z$ is the $N$-vector of subject-level treatments $Z_i$ and $W$ is the matrix of edge-level treatments $W_{ij}$. The marginal distribution is thus a function that maps a subject ($j$) or edge ($ij$) to probability of treatment.  More advanced experiments might additionally allow this function to depend on pre-treatment covariates $X_j$ or the existing interaction network $A_{ij}$. The specific form of the randomization determines what causal questions the experiment is capable of, or especially suitable for, answering.

\subsubsection{Implementing randomization}
\label{sec:implementing}
In practice, researchers tend to implement randomization using deterministic cryptographic hash functions to generate pseudo-random variables with specified distributions \citep{kohavi2012trustworthy,bakshy_designing_2014}. PlanOut is a domain-specific language for specifying randomization strategies that is used at Facebook and several other companies.\footnote{The design of PlanOut is described in \citet{bakshy_designing_2014} and it is available from \href{https://github.com/facebook/planout}{https://github.com/facebook/planout}.}  Using variable-specific crypographic salts, PlanOut provides functionality for independent random assignment for multiple experiments, multiple variables, and multiple types of units (e.g., users, clusters, items, edges). The determinism of the hash functions ensures that a random assignment is ``persistent'', without requiring the assignments be stored; that is, the assignments can be computed online and statelessly, as subjects arrive. PlanOut code implementing the i.i.d. randomization we described in the previous paragraph as well as some more advanced randomizations are shown in Listing~\ref{lst:planout_iid}.

\begin{lstlisting}[caption={Example PlanOut code for subject-level treatment assignment.}, label={lst:planout_iid}] 
# i.i.d. random assignment
smoking_program = uniformChoice(choices=[0,1], unit=subject_id);

# block random assignment
smoking_program = uniformChoice(choices=[0,1], unit=subject_group_id);

# hierarchical block random assignment
smoking_program_prob = randomFloat(min=0, max=1, unit=subject_group_id);
smoking_program = bernoulliTrial(p=smoking_program_prob, unit=subject_id);
\end{lstlisting}

\subsubsection{Subject-level treatment randomizations} \label{sec:subject_randomization}

\begin{figure}[t]
\label{fig:randomization_diagram}
\centering
\begin{tikzpicture}
 
\begin{scope}[scale = 0.75]

\draw[anchor = west, text width = 4cm] (-0.5,5) node {i.i.d. assignment on \\ balanced groups};

% grid
\foreach \s in {0,1,2,3,4}{
  \draw[gray] (0,\s) -- (4,\s);
  \draw[gray] (\s,0) -- (\s,4);
}
\fill[black, opacity = 0.5] (0,4) rectangle (4,3);
\fill[black, opacity = 0.5] (0,2) rectangle (4,1);

\foreach \r in {0,1,2,3} {
  \foreach \x in {0,1,2,3} {
    \foreach \y in {0,1,2,3} {
      \fill[black] (\r + 0.125 + \x*0.25, 4 - \r - 0.125 - \y*0.25) circle [radius = 0.05];
    }
  }
}

\draw[gray] (0, -4) -- (4, -4);
\fill[black, opacity = 0.5] (0.25,-4) rectangle (0.75,-1);
\fill[black, opacity = 0.5] (3.25,-4) rectangle (3.75,-1);

\draw (0.50,-4.5) node {0};
\draw (3.50,-4.5) node {1};

\end{scope}

\begin{scope}[xshift=4cm, scale = 0.75]

\draw[anchor=west, text width=4cm] (-0.5,5) node {hierarchical assignment \\ on balanced groups};

\foreach \s in {0,1,2,3,4}{
  \draw[gray] (0,\s) -- (4,\s);
  \draw[gray] (\s,0) -- (\s,4);
}

\fill[black, opacity = 0.5] (0,4) rectangle (4,3);
\fill[black, opacity = 0.3] (0,3) rectangle (4,2);
\fill[black, opacity = 0.1] (0,2) rectangle (4,1);

\foreach \r in {0,1,2,3} {
  \foreach \x in {0,1,2,3} {
    \foreach \y in {0,1,2,3} {
      \fill[black] (\r + 0.125 + \x*0.25, 4 - \r - 0.125 - \y*0.25) circle [radius = 0.05];
    }
  }
}

\draw[gray] (0, -4) -- (4, -4);
\fill[black, opacity = 0.5] (0.25,-4) rectangle (0.75,-2.5);
\fill[black, opacity = 0.5] (1.25,-4) rectangle (1.75,-2.5);
\fill[black, opacity = 0.5] (2.25,-4) rectangle (2.75,-2.5);
\fill[black, opacity = 0.5] (3.25,-4) rectangle (3.75,-2.5);

\draw (0.50,-4.5) node {0};
%\draw (1.50,-4.5) node {.33};
%\draw (2.50,-4.5) node {.66};
\draw (3.50,-4.5) node {1};

\end{scope}

\begin{scope}[xshift=8cm, scale = 0.75]

\draw[anchor=west, text width=4cm] (-0.5,5) node {i.i.d.  assignment on \\ small world network};

\draw[gray] (0,0) -- (4,0);
\draw[gray] (0,4) -- (4,4);
\foreach \s in {0,1,2,3,4}{
  \draw[gray] (0,\s) -- (4,\s);
  \draw[gray] (\s,0) -- (\s,4);
}

\fill[black, opacity = 0.5] (0,4) rectangle (4,2);

\fill[black] (0.125,3.625) circle [radius = .05];
\fill[black] (0.375,3.375) circle [radius = .05];
\fill[black] (0.625,3.125) circle [radius = .05];
\fill[black] (1.625,0.125) circle [radius = .05];
\fill[black] (1.375,1.625) circle [radius = .05];
\fill[black] (1.375,2.375) circle [radius = .05];
\fill[black] (0.125,2.375) circle [radius = .05];
\fill[black] (1.125,2.125) circle [radius = .05];
\fill[black] (0.625,1.875) circle [radius = .05];
\fill[black] (0.625,1.375) circle [radius = .05];
\fill[black] (2.625,1.125) circle [radius = .05];
\fill[black] (2.875,0.875) circle [radius = .05];
\fill[black] (3.125,0.625) circle [radius = .05];
\fill[black] (1.625,0.375) circle [radius = .05];
\fill[black] (1.875,0.125) circle [radius = .05];
\fill[black] (0.125,2.125) circle [radius = .05];
\fill[black] (0.125,3.375) circle [radius = .05];
\fill[black] (0.625,0.375) circle [radius = .05];
\fill[black] (0.375,0.375) circle [radius = .05];
\fill[black] (0.375,3.125) circle [radius = .05];
\fill[black] (0.625,0.125) circle [radius = .05];
\fill[black] (0.875,2.625) circle [radius = .05];
\fill[black] (1.625,1.375) circle [radius = .05];
\fill[black] (1.375,2.125) circle [radius = .05];
\fill[black] (1.625,1.875) circle [radius = .05];
\fill[black] (1.875,1.625) circle [radius = .05];
\fill[black] (2.125,0.375) circle [radius = .05];
\fill[black] (2.375,1.125) circle [radius = .05];
\fill[black] (2.625,0.875) circle [radius = .05];
\fill[black] (2.375,0.625) circle [radius = .05];
\fill[black] (3.125,0.375) circle [radius = .05];
\fill[black] (0.625,0.625) circle [radius = .05];
\fill[black] (0.375,3.875) circle [radius = .05];
\fill[black] (0.625,3.625) circle [radius = .05];
\fill[black] (0.875,3.375) circle [radius = .05];
\fill[black] (3.875,2.375) circle [radius = .05];
\fill[black] (2.375,2.625) circle [radius = .05];
\fill[black] (1.625,2.625) circle [radius = .05];
\fill[black] (1.625,3.875) circle [radius = .05];
\fill[black] (1.875,2.875) circle [radius = .05];
\fill[black] (2.125,3.375) circle [radius = .05];
\fill[black] (2.625,3.375) circle [radius = .05];
\fill[black] (2.875,1.375) circle [radius = .05];
\fill[black] (3.125,1.125) circle [radius = .05];
\fill[black] (3.375,0.875) circle [radius = .05];
\fill[black] (3.625,2.375) circle [radius = .05];
\fill[black] (3.875,2.125) circle [radius = .05];
\fill[black] (1.875,3.875) circle [radius = .05];
\fill[black] (0.625,3.875) circle [radius = .05];
\fill[black] (3.625,3.375) circle [radius = .05];
\fill[black] (3.625,3.625) circle [radius = .05];
\fill[black] (0.875,3.625) circle [radius = .05];
\fill[black] (3.875,3.375) circle [radius = .05];
\fill[black] (1.375,3.125) circle [radius = .05];
\fill[black] (2.625,2.375) circle [radius = .05];
\fill[black] (1.875,2.625) circle [radius = .05];
\fill[black] (2.125,2.375) circle [radius = .05];
\fill[black] (2.375,2.125) circle [radius = .05];
\fill[black] (3.625,1.875) circle [radius = .05];
\fill[black] (2.875,1.625) circle [radius = .05];
\fill[black] (3.125,1.375) circle [radius = .05];
\fill[black] (3.375,1.625) circle [radius = .05];
\fill[black] (3.625,0.875) circle [radius = .05];
\fill[black] (3.375,3.375) circle [radius = .05];
\fill[black] (0.125,3.875) circle [radius = .05];
\fill[black] (0.375,3.625) circle [radius = .05];
\fill[black] (0.625,3.375) circle [radius = .05];
\fill[black] (0.875,3.125) circle [radius = .05];
\fill[black] (1.125,2.875) circle [radius = .05];
\fill[black] (1.375,2.625) circle [radius = .05];
\fill[black] (1.625,2.375) circle [radius = .05];
\fill[black] (1.875,2.125) circle [radius = .05];
\fill[black] (2.125,1.875) circle [radius = .05];
\fill[black] (2.375,1.625) circle [radius = .05];
\fill[black] (2.625,1.375) circle [radius = .05];
\fill[black] (2.875,1.125) circle [radius = .05];
\fill[black] (3.125,0.875) circle [radius = .05];
\fill[black] (3.375,0.625) circle [radius = .05];
\fill[black] (3.625,0.375) circle [radius = .05];
\fill[black] (3.875,0.125) circle [radius = .05];

\draw[gray] (0, -4) -- (4, -4);
% -3.8125 -3.0625 -2.1250 -2.1250 -3.0625
\fill[black, opacity = 0.5] (0.25,-4) rectangle (0.75,-3.8125);
\fill[black, opacity = 0.5] (1.00,-4) rectangle (1.50,-3.0625);
\fill[black, opacity = 0.5] (1.75,-4) rectangle (2.25,-2.1250);
\fill[black, opacity = 0.5] (2.50,-4) rectangle (3.00,-3.0625);
\fill[black, opacity = 0.5] (3.25,-4) rectangle (3.75,-3.8125);

\draw (0.50,-4.5) node {0};
%\draw (1.25,-4.5) node {.25};
%\draw (2.00,-4.5) node {.5};
%\draw (2.75,-4.5) node {.75};
\draw (3.50,-4.5) node {1};

\end{scope}

\begin{scope}[xshift=12cm, scale = 0.75]

\draw[anchor=west, text width=4cm] (-0.5,5) node {clustered  assignment on \\ small world network};

\draw[gray] (0,0) -- (4,0);
\draw[gray] (0,4) -- (4,4);
\foreach \s in {0,1,2,3,4}{
  \draw[gray] (0,\s) -- (4,\s);
  \draw[gray] (\s,0) -- (\s,4);
}

\fill[black, opacity = 0.5] (0,4) rectangle (4,2);

\fill[black] (0.125,3.625) circle [radius = .05];
\fill[black] (0.375,3.375) circle [radius = .05];
\fill[black] (0.625,3.125) circle [radius = .05];
\fill[black] (1.125,2.125) circle [radius = .05];
\fill[black] (2.375,1.125) circle [radius = .05];
\fill[black] (2.375,2.875) circle [radius = .05];
\fill[black] (0.125,2.875) circle [radius = .05];
\fill[black] (2.125,1.375) circle [radius = .05];
\fill[black] (0.625,2.625) circle [radius = .05];
\fill[black] (0.625,0.875) circle [radius = .05];
\fill[black] (3.125,0.625) circle [radius = .05];
\fill[black] (3.375,0.375) circle [radius = .05];
\fill[black] (3.625,0.125) circle [radius = .05];
\fill[black] (1.125,2.375) circle [radius = .05];
\fill[black] (2.625,2.125) circle [radius = .05];
\fill[black] (0.125,1.375) circle [radius = .05];
\fill[black] (0.125,3.375) circle [radius = .05];
\fill[black] (0.625,2.375) circle [radius = .05];
\fill[black] (0.375,2.375) circle [radius = .05];
\fill[black] (0.375,3.125) circle [radius = .05];
\fill[black] (0.625,2.125) circle [radius = .05];
\fill[black] (0.875,1.625) circle [radius = .05];
\fill[black] (1.125,0.875) circle [radius = .05];
\fill[black] (2.375,1.375) circle [radius = .05];
\fill[black] (1.125,2.625) circle [radius = .05];
\fill[black] (2.625,1.125) circle [radius = .05];
\fill[black] (1.375,2.375) circle [radius = .05];
\fill[black] (2.875,0.625) circle [radius = .05];
\fill[black] (3.125,0.375) circle [radius = .05];
\fill[black] (2.875,0.125) circle [radius = .05];
\fill[black] (3.625,2.375) circle [radius = .05];
\fill[black] (0.625,0.125) circle [radius = .05];
\fill[black] (0.375,3.875) circle [radius = .05];
\fill[black] (0.625,3.625) circle [radius = .05];
\fill[black] (0.875,3.375) circle [radius = .05];
\fill[black] (1.875,2.875) circle [radius = .05];
\fill[black] (2.875,1.625) circle [radius = .05];
\fill[black] (1.125,1.625) circle [radius = .05];
\fill[black] (1.125,3.875) circle [radius = .05];
\fill[black] (2.625,1.875) circle [radius = .05];
\fill[black] (1.375,3.375) circle [radius = .05];
\fill[black] (3.125,3.375) circle [radius = .05];
\fill[black] (3.375,0.875) circle [radius = .05];
\fill[black] (3.625,0.625) circle [radius = .05];
\fill[black] (3.875,0.375) circle [radius = .05];
\fill[black] (1.625,2.875) circle [radius = .05];
\fill[black] (1.875,1.375) circle [radius = .05];
\fill[black] (2.625,3.875) circle [radius = .05];
\fill[black] (0.625,3.875) circle [radius = .05];
\fill[black] (1.625,3.375) circle [radius = .05];
\fill[black] (1.625,3.625) circle [radius = .05];
\fill[black] (0.875,3.625) circle [radius = .05];
\fill[black] (1.875,3.375) circle [radius = .05];
\fill[black] (2.375,3.125) circle [radius = .05];
\fill[black] (3.125,2.875) circle [radius = .05];
\fill[black] (2.625,1.625) circle [radius = .05];
\fill[black] (1.375,2.875) circle [radius = .05];
\fill[black] (2.875,1.375) circle [radius = .05];
\fill[black] (1.625,2.625) circle [radius = .05];
\fill[black] (3.375,1.125) circle [radius = .05];
\fill[black] (3.625,0.875) circle [radius = .05];
\fill[black] (3.875,1.125) circle [radius = .05];
\fill[black] (1.625,0.375) circle [radius = .05];
\fill[black] (3.875,3.375) circle [radius = .05];
\fill[black] (0.125,3.875) circle [radius = .05];
\fill[black] (0.375,3.625) circle [radius = .05];
\fill[black] (0.625,3.375) circle [radius = .05];
\fill[black] (0.875,3.125) circle [radius = .05];
\fill[black] (2.125,1.875) circle [radius = .05];
\fill[black] (2.375,1.625) circle [radius = .05];
\fill[black] (1.125,2.875) circle [radius = .05];
\fill[black] (2.625,1.375) circle [radius = .05];
\fill[black] (1.375,2.625) circle [radius = .05];
\fill[black] (2.875,1.125) circle [radius = .05];
\fill[black] (3.125,0.875) circle [radius = .05];
\fill[black] (3.375,0.625) circle [radius = .05];
\fill[black] (3.625,0.375) circle [radius = .05];
\fill[black] (3.875,0.125) circle [radius = .05];
\fill[black] (1.625,2.375) circle [radius = .05];
\fill[black] (1.875,2.125) circle [radius = .05];

\draw[gray] (0, -4) -- (4, -4);
% -3.2125 -2.6500 -2.2750 -2.6500 -3.2125
\fill[black, opacity = 0.5] (0.25,-4) rectangle (0.75,-3.2125);
\fill[black, opacity = 0.5] (1.00,-4) rectangle (1.50,-2.6500);
\fill[black, opacity = 0.5] (1.75,-4) rectangle (2.25,-2.2750);
\fill[black, opacity = 0.5] (2.50,-4) rectangle (3.00,-2.6500);
\fill[black, opacity = 0.5] (3.25,-4) rectangle (3.75,-3.2125);

\draw (0.50,-4.5) node {0};
%\draw (1.25,-4.5) node {.25};
%\draw (2.00,-4.5) node {.5};
%\draw (2.75,-4.5) node {.75};
\draw (3.50,-4.5) node {1};

\end{scope}

\end{tikzpicture}
\caption{Various subject-level randomizations illustrating how they each induce different distributions of treatment status for a subject's friends.  Each of the four squares is an adjacency matrix (dots represent undirected friendships).  The horizontal grey bars represent treatment probabilities, with the darkest color indicating treatment is assigned to subjects in that row with 100\% probability.  The stylized histograms beneath the squares indicate the fraction of friends who are treated induced by the randomization strategy above it.}
\end{figure}
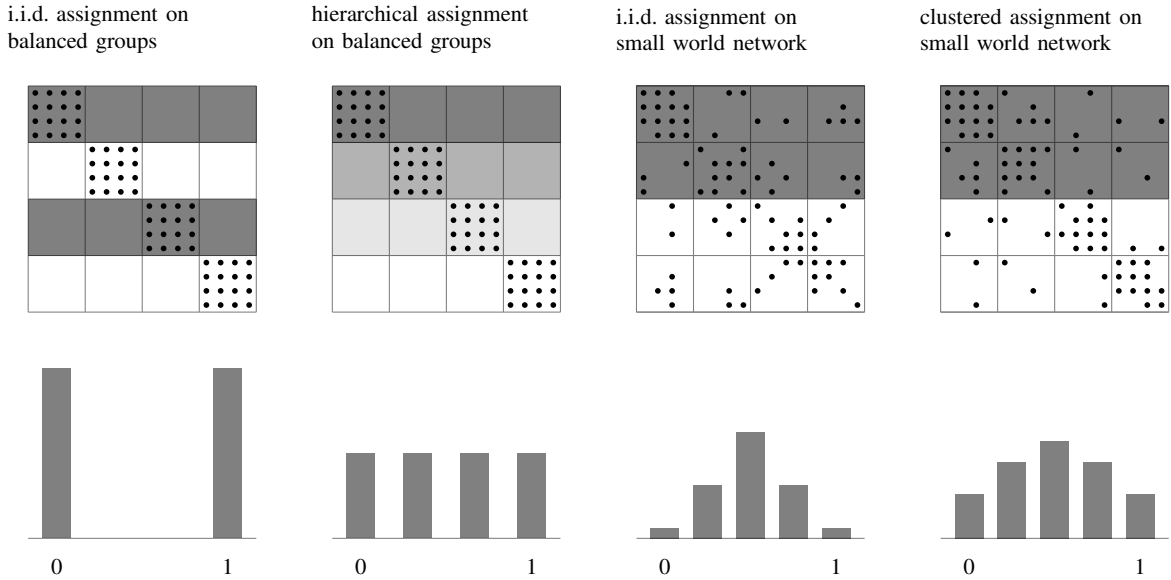
Here we consider randomizations for subject-level treatments. Consider the simplest possible randomization for a subject-level treatment is independent and identically distributed (i.i.d) Bernoulli random variable: $Z_j \sim \mbox{Bernoulli}(0.5)$.\footnote{Often the literature on randomized experiments \citep[e.g.,][]{gerber2012field,imbens2015causal} starts with a \emph{completely randomized design}, in which some fixed number $N_1$ of the $N$ subjects are assigned to treatment. However, in the case of large digital field experiments implemented as described in Section \ref{sec:implementing}, this cannot easily be done in online (i.e. streaming) assignment without complications.}
Say $Z_j$ is assignment to a smoking prevention program. We hypothesize that the program will reduce how much people in the study smoke (i.e. $D_j$ is lower in expectation when $Z_j = 1$), and are further interested in using this randomization to learn about social influence in smoking. In the context of disjoint groups (i.e. a network consisting of multiple disjoint cliques), we can think of this randomization as a \emph{partial population experiment} \citep{moffitt2001policy,baird2016optimal}, in that some of the population is treated and we can study behavior of their peers. This design is analogous to marketing interventions which seek to exploit spillovers or network effects in demand by providing discounts or promotions to a small subset of consumers \citep{hartmann2010demand}.

In order for our randomization to enable detecting and estimating social influence, we will generally need variation in the treatments of the peers of our subjects. While many measures of peer treatment can be used, we will illustrate the points in this section with the fraction of $i$'s peers who are treated:
$$
T_i = \sum_{j=1}^N \bar{A}_{ij} Z_j,
$$
where $\bar{A}_{ij} = A_{ij} / \sum_{j=1}^N A_{ij}$ an entry in the row-normalized adjacency matrix, with $\bar{A}_{ij} = 0$ if $\sum_{j=1}^N A_{ij} = 0$.

The i.i.d. subject-level assignment described above and shown in the third panel of Figure~\ref{fig:randomization_diagram} has a very important limitation: if a subject has a substantial number of peers, then there is a vanishingly small probability that they will all be assigned to treatment; for example, if subject $i$ has 10 peers then $\Pr(T_i = 1) = \Pr(\sum_{i=1}^{10} Z_j = 10) < .01$.  So we are unlikely to be able to use an experiment with this type of randomization to answer counterfactual questions about having all (or even a large percentage) of a person's friends participate in the program. For some asymptotic sequences with growing degree, this will mean the variance of sample means for units with, e.g., all treated peers diverges \citep{ugander_graph_2013}.  Thus, we will often want to consider other randomizations.

At the opposite extreme, we could assign treatment at the level of groups or clusters.
For instance if students are grouped by classrooms, we could do the smoking prevention assignment at the classroom-level.  Let $c(j)$ be the classroom for subject $j$. Then a group-level randomization would be to assign each group an i.i.d Bernoulli, $P_c \sim \mbox{Bernoulli}(0.5)$ and assign each student her group's assignment, $Z_j = P_{c(j)}$.
If we think the classrooms are disjoint cliques, we might posit an interaction network that is a block-diagonal matrix, such that $A_{ij} = \mathbbm{1}\{c(i) = c(j)\}$.
Note that in the case of disjoint groups, such an ``everyone or nobody'' randomization abandons the partial population idea. This randomization can help answer questions about what will happen should we deploy the program to everyone, but it cannot answer questions about social influence and thus whether the program can be deployed more cost-effectively by treating a smaller proportion of students.  We may be able to dramatically reduce smoking in a classroom by encouraging 25\% of the students to not smoke.  In this group randomization, we will never observe a classroom with any quantity other than 0\% or 100\% of the students treated; see the first panel of Figure~\ref{fig:randomization_diagram}.

Intermediate designs between these two extremes use a hierarchical (or, in this case, two-stage) randomization to creates additional dispersion in the quantity of students per classroom assigned to the treatment, but also makes subject's own treatment and their peers not perfectly dependent. For example, we can first draw a random uniform variable per classroom, $P_g \sim \mbox{Uniform}(0, 1)$, and then for each student, we draw a Bernoulli random variable with their group's probability, $Z_j \sim \mbox{Bernoulli}(p_{c(j)})$.\footnote{
With a small number of groups, we may want to use a completely randomized design, rather than independent draws of $P_c$
\citet{baird2016optimal} consider optimal two-stage randomizations in the context of disjoint groups, given the goal of estimating some particular direct or indirect effects.
}
For some randomization $\psi$, we call it \emph{overdispersed} because $\Var_\psi(T_i) > \Var_{\text{iid}}(T_i)$; that is, it greater variability in the fraction of peers treated than from i.i.d. subject-level randomizations.
An overdispersed randomization could be useful for selecting a number of students to treat per classroom, given some budget, that will minimize smoking because it can provide an estimate smoking behavior under different many levels of treatment.  %\cite{duflo_role_2003} use a similar randomization in order to test whether decisions to enroll in a retirement plan are affected by social learning.

Block-diagonal  networks (e.g., villages assumed to not interact) make overdispersed randomizations easy to implement. With more general networks, there are more design choices, and it can be difficult to generate arbitrary degrees of overdispersion in friend treatment assignment probabilities.
We would like a randomization such that the distribution of $T_i$ has certain properties; for example, one heuristic, is we should have that $\Pr(T_i = k) > \varepsilon$ for all feasible fractions $k$ given $i$'s degree. Or we may aim to maximize $\Pr(T_i = k)$ for $k \in \{0, 1\}$.
One recently popular way to do so is to partition the network into clusters using existing graph partitioning algorithms, and then proceed with the cluster-randomized design \citep[i.e. \emph{graph cluster randomization;}][]{ugander_graph_2013,eckles_design_2017,xu2015infrastructure}.
Given the structure of the network, there will still be edges between clusters (fourth panel in Figure \ref{fig:randomization_diagram}). For example, say we use state-of-the art methods to partitioning the Facebook friendship network; with only 1,000 clusters, already over 40\% of edges will be between clusters \citep{shalita2016social}.
Not only is graph partitioning challenging in large networks, but standard min-cut objectives will often just be a heuristic: we would instead prefer to optimize bias or total error in estimation of particular quantities. 
%For graph cluster randomizations, there is a large set of graph partitioning techniques the researcher can choose, but perhaps more important is into how many clusters the network should be partitioned. If the researcher generates large number of small clusters, the result is lower variance estimation because there are more ``units'' to analyze. However, large clusters yield less biased estimates because they retain more similar graph structure to the whole graph.
To facilitate such optimization, one can further treat the clusters, or some other model fit to the network (e.g., a more general stochastic block model \citep{karrer_stochastic_2011}), as an approximation to the observed network. Thus, \citet{basse2015optimal} propose using optimal designs for approximations to the observed network.

A final design possibility with subject-level treatment randomizations is that treatment assignment probabilities can depend on pre-treatment covariates $X_i$ in order to increase precision. \emph{Blocking} or \emph{pre-stratification} exactly balances some covariates between treatments, rather than simply balancing them in expectation, thus reducing the variance in effect estimates is attributable to the random assignment of treatments causing covariate imbalance in small samples \citep[ch. 4]{gerber2012field}.  For instance, in a small sample it could make a large difference in estimates if a subject who is very active or who has many friends is assigned to treatment or not.  State-of-the-art blocking methods allow improving balancing on high-dimensional covariates and lead to higher-precision estimates of treatment effects \citep{higgins2016improving}. While usually large samples make blocking irrelevant because post-stratification or regression adjustment can provide similar precision gains \citep{miratrix2013adjusting}, use of graph cluster randomization again reduces the effective number of units being randomized, perhaps making blocking a relevant design consideration.

Pre-treatment covariates can be used to target specific subjects who may have certain network positions or be likelier to cause social influence based on some hypothesis or prior analysis.  If a researcher wanted to test a seeding strategy based on network position a reasonable design would be select a set of influential candidate subjects \citep{kempe2003maximizing} and treat a random fraction of them while reserving some others as a control \citep[cf.][]{kim2015social, beaman_can_2015}.

\subsubsection{Interaction-level treatment randomization}
Treatments defined at the level of individual edges allow for further choices in randomization. Because this design space is so large, we consider some notable examples.

Historically, many examples of interaction-level treatments come from experiments in the formation of random groups. Here the interaction network is set in advance by the researcher or by some exogenous process.  From a notational standpoint, these designs amount to setting $W_{ij} = 1$ for blocks of subjects to induce variation in $A_{ij}$ and, in turn, the distribution of quantities such as the fraction of adopting peers, $\sum_{j=1}^N D_j \bar{A}_{ij}$.  An important aspect of this type of randomization is that the resulting groups must exhibit variance on $D_j$, the behavior of interest.  For the same reason that i.i.d. assignment in subject-level treatments may not cause sufficient variation in peer exposures, large random groups are unlikely to be useful for identifying causal effects \citep[cf.][]{angrist_perils_2014}. As with subject-level treatments above, it may be desirable to introduce overdispersion in group composition.

The random group assignment designs generally leverage existing group formation policies. In the case of \cite{sacerdote2001peer}, which exploits the fact that roommate assignments at Dartmouth college are conditionally randomly assigned (directly setting $A_{ij} = 1$ for the ``is roommate'' relation), we may even consider this a natural experiment. %Using this existing processes which induces pairs of students who are in prolonged contact with person whose ability, motivation, and interests are completely independent of their own.  
On the other hand, \cite{carrell_does_2009,carrell_natural_2011} introduce novel group formation policies for squadrons at the United States Air Force Academy; here squadrons are groups of roughly 30 that cadets are required to spend the majority of their time with. %\cite{carrell_does_2009} and \cite{sacerdote2001peer} are notable for employing a very strong treatment and therefore being capable of generating measurable effects that are economically significant.
As a further refinement, random group formation can be performed dynamically to allow for repeated measurements of the same individuals as they change social contexts.  \cite{hasan2017conversational} use such a randomization to measure how group interactions between entrepreneurs affects their ideation.  Their approach allows them to not only measure how changing groups affects their outcome of interest, but allows for longitudinal measurements of individual outcomes as well.

Without leveraging existing group formation policies, researchers may be limited to encouraging the formation edges that involve less prolonged contact. Several experiments have randomly assigned subjects to different graph structures whether in an artificial setting \cite[e.g.,][]{kearns2006experimental} or in the context of an online health-related service \citep{centola2010spread}. Here the experiment is generally conceptualized at the level of entire replications of a particular graph. Thus, the outcomes and analyses may be defined and conducted in aggregate rather than at the individual level.  One can think of these designs as randomizing $A$ directly and then observing some aggregate network outcome, which is slightly more complex than the framework we propose here.

Other edge-level treatments are best understood as conditional on peer behaviors and a pre-treatment network. These include what we have called mechanism experimental designs, which work by randomizing whether a social signal is delivered via particular channel. Mechanism designs \citep[e.g.,][]{bakshy_role_2011,aral2012identifying,bond_experiment_2012} are equipped to answer counterfactuals about how peer behavior would be affected in the amplification or attenuation of the influence channel of interest.\footnote{An additional refinement of the model we outline here is subjects may be connected via multiple overlapping networks, such as in-person vs online interactions, and an experiment may cause changes in some of those networks but not others.} 
For example, in \cite{bakshy_social_2012}, the only peers eligible for the experiment are those who have already liked a page on Facebook (conditioning on $D_j = 1$) and the randomization (assigning $W_{ij}$ as a Bernoulli random variable with perfect compliance for $A_{ij}$) determines whether this behavior will be displayed when the focal user sees an ad.
\cite{aral2012identifying} use another mechanism design in exploiting the fact that notifications in their Facebook application are delivered to a random set of the user's friends.  If we believe that these notifications are the only mechanism through which a Facebook friend might adopt the application, this amounts to randomly amplifying values of $A_{ij}$ for the friends who received the notifications, while leaving it un-amplified for the remaining Facebook friends that were collected when the user installed the application.

Just because a treatment is defined at the level of edges not mean the randomization is i.i.d at the edge level. \cite{bakshy_social_2012} select random subsets of edges involving the same subject. In the context of a treatment that encourages providing feedback (likes and comments on Facebook, in this case) to a specified directed edge, \citet{eckles_estimating_2016} compare different possible randomizations. One sender-clustered design would randomly assign vertices to an encouragement to give all of their peers more feedback. Another recipient-clustered design would randomly assign vertices to have \emph{all of their peers} encouraged to give them feedback This latter design is used in \citet{eckles_estimating_2016}, as simulations suggest it will often have precision advantages.
Finally, other designs could, like some of the designs we considered in the previous section, interpolate between i.i.d. assignment of edges and either of these clustered designs.

\subsection{Outcome measurement} \label{sec:outcomes}

Perhaps an underrated requirement of randomized experiments is the ability to measure an outcome appropriate to the research question at hand.  Sometimes researchers invest more time and expense in intervening with their treatment than in measuring the outcome.  However, precise, valid, and complete measurement plays a large role in the success of randomized experiments. 

A simple example is that, if outcomes are measured with noise, the resulting estimates will be less precise. Even more problematic are cases where some outcomes are missing, either randomly or not.  \cite{coey2016people} show that matching ad exposures to conversions via cookies --- where matching is random but plausibly independent of treatment status --- results in a substantial loss of experimental power.  Other experiments might rely on surveys or self-reports to measure outcomes, which yields either a biased measurement (e.g., social desirability) or a treatment effect estimate for only a biased sub-population (survey takers).  \cite{berry2017discussion}, who study social influence for comment quality in public discussions, can only measure comment quality improvements for the set of subjects who choose to write comments.  \cite{bond_experiment_2012} measure voter turnout by matching Facebook users to people in the state voter files, which is a noisy process (match rates were about 40\%) that was limited to 13 states because of the expense of acquiring voter file data.

Digital field experiments present some opportunities and also limitations for experimenters.  Many important outcomes are potentially observable, such as clicks on ads \cite{bakshy_social_2012}, sharing and production of user-generated content \citep{bakshy_role_2011,eckles_estimating_2016}, and adoption of apps (both free and paid) \citep{aral2011creating,bapna2015your}.  However, digital platforms create comprehensive logs of \emph{digital} behaviors, which are perhaps not the only behaviors of theoretical interest.  For instance, while \cite{kramer2014experimental} apply a reasonable text-analysis procedure to measure people's emotions at scale, it is debatable whether a change in emotion is adequately captured by the text they choose to share on Facebook \citep{beasley2015emotional}.  The sheer volume of data produced on digital platforms is a signal of how trivial the actions they collect can be.  Despite dramatic advances in observability of human behavior, it continues to be a central research challenge to measure important outcomes and join them to experimentally assigned treatments.

\section{Analyzing randomized experiments} \label{sec:analyzing}

One frequent consequence of having a well-designed randomized experiment is that the data analysis is then straightforward. While this is true to some degree in experiments about social influence in networks, estimation and inference can both be complicated by the network. Causal and statistical inference in networks remains an active research area, with contemporary contributions to basic problems such as laws of large numbers and asymptotic inference in networks \citep{leung_weak_2017,aronow2013estimating,tchetgen2017auto,laan_causal_2014}.

In this section, we review methods for estimation and inference (e.g., hypothesis testing) for social influence in network experiments. The known randomization of subjects or edges to treatments provides a ``reasoned basis'' for inference \citep[p. 14]{fisher1935design} with minimal assumptions even when we only observe a network with a single giant component. We thus focus on Fisherian randomization inference, but briefly review other methods.

As with the experimental design, the primary goal in analysis is learning about social influence. Ideally, this means learning about effects of $D_j$ on $Y_i$ or of $A_{ij}$ on $Y_i$. It will often be more straightforward to simply detect any effects of $Z_j$ or $W_{ij}$ on $Y_i$. This is because (a) the experimenter sets these, but usually only affects $D_j$ on $Y_i$ indirectly and (b) in measuring $D_j$ and $A_{ij}$, we may not capture all of the ways that our treatments can affect subjects. Thus, we can often take evidence about effects of $Z_j$ or $W_{ij}$ as evidence of social influence, without being about to denominate these effects in terms of peer behaviors. We start with this simpler case.

\subsection{Effects of randomized treatments}
In this section, we consider how to conduct inference about effects of randomized treatments. We start by considering inference about spillovers in experiments where subjects are randomly assigned to subject-level treatments; that is, we are interested in questions about whether subjects' outcomes are affected by others' treatments. If we assume that others' treatment only affect an individual through others' behaviors (Figure \ref{fig:diagram}), then these tests are also tests of social influence.

\subsubsection{Testing sharp null hypotheses about spillovers}

Consider the null model in which there is a direct effect of a subject's own treatment, but no effects of others' treatments, including those of peers.
\begin{hypothesis}[No spillovers with constant direct effects]
\label{hyp:constant_direct_effect}
There exists some $\tau$ such that $Y_i(z_i) = \tau z_i + \xi_i$ for all $z \in \zbm^N$ and $i \in V$.
\end{hypothesis}
Note that under this null hypothesis $Y_i - \tau Z_i$ does not vary under alternative treatment assignments. This null hypothesis is a composite of null hypotheses of the form:
\begin{hypothesis}[No spillovers with constant direct effects, $\tau_0$]
\label{hyp:constant_direct_effect_sharp}
$Y_i(z_i) = \tau_0 z_i + \xi_i$ for all $z \in \zbm^N$ and $i \in V$.
\end{hypothesis}
Hypothesis \ref{hyp:constant_direct_effect_sharp} is a \emph{sharp null hypothesis}, which allows inferring all of a unit's potential outcomes from its single, observed potential outcome.
We can thus use Fisherian randomization inference, in which we exploit our knowledge of the distribution of $Z$ (which we or the experimenter chose), to test this null hypothesis. This is often implemented as a permutation test with a test statistic chosen to be sensitive to the kinds of deviations from the null that we expect. For example, consider a larger model that includes a linear effect of the fraction of treated peers:
\begin{equation}
\label{eq:spillovers_frac}
Y_i = \tau Z_i + \rho \sum_{j = 1}^N Z_j \bar{A}_{ij} + \xi_i 
\end{equation}
where $\bar{A}_{ij}$ is an entry in the row normalized adjacency matrix. A non-zero $\rho$ would correspond to a particular violation of Hypothesis \ref{hyp:constant_direct_effect}. The score statistic for $\rho$ can be used as a test statistic \citep{athey_exact_2017}, as can many other test statistics.

Algorithm \ref{algo:sharp} tests Hypothesis \ref{hyp:constant_direct_effect_sharp} using Fisherian randomization inference. To test Hypothesis \ref{hyp:constant_direct_effect}, researchers would generally test many particular values of $\tau$ (e.g., in a grid, or through a search algorithm) and take the supremum.

\begin{algo}[Randomization inference for Hypothesis \ref{hyp:constant_direct_effect_sharp}]
\label{algo:sharp}
Inputs: test statistic $T(\cdot, \cdot) : \mathbb{Y}^N \times \{0, 1\}^N$ that is a function of units’ residual outcomes and the treatment vector; posited direct effect $\tau_0$.
\begin{enumerate}
\item Compute residual outcomes given $\tau_0$, $\tilde{Y} := Y - \tau_0 Z$.
\item For every $r \in \{1, ..., R\}$ and some $\tau_0$:
\begin{enumerate}
\item Draw a new treatment vector $Z^*$ consistent with the original randomization.
\item Compute value of test statistic with observed outcomes and permuted treatment $T_{\text{null}, r} := T(\tilde{Y}, Z^*)$.
\end{enumerate}
\item Compare observed and null test statistics, yielding
$$\widehat{\text{p-value}}(\tau_0) = \frac{1}{R} \sum_{r = 1}^R \mathbbm{1} \{T(\tilde{Y}, Z) > T_{\text{null}, r} \}.$$
\end{enumerate}
\end{algo}

We would then reject Hypothesis \ref{hyp:constant_direct_effect_sharp} for small p-values, instead concluding subjects are affected by others' treatments.

\begin{remark}[Randomization inference and permutation tests]
While randomization inference frequently makes use of permutation tests, the two are not identical. Fisherian randomization inference makes use of knowledge about the exact distribution of variables that were randomized to conduct exact causal inference for a finite population of units. Often (e.g. with a single completely randomized treatment vector) this can be approximated to arbitrary precision through permutation of the treatment vector, but need not be if the distribution over treatments is more complicated. Furthermore, permutation tests of social influence are often used without the justification they are afforded by randomization; that is, they are often used when other assumptions would be needed to make them exact in finite samples or even good asymptotic approximations. For example, \citet{anagnostopoulos2008influence} make additional, strong assumptions about non-influence processes to justify the use of a permutation test to detect influence in observational data.

Even in the case of randomized experiments, particular permutation tests may not be readily justified by the randomization. Without explicitly considering the relevant sharp null hypothesis, it can be easy to make mistakes that make the resulting permutation test invalid. For example, \citet{bond_experiment_2012} test for spillovers from a randomly assigned encouragement to vote in the 2010 U.S. elections. This was implemented as as a permutation test that implicitly assumed the absence of direct effects, even though \citet{bond_experiment_2012} elsewhere rejected that null hypothesis. \citet{athey_exact_2017} show that such tests can have dramatically inflated Type I error rates (i.e. they too often reject the null hypothesis when it is true).
\end{remark}

\subsubsection{Inference for the magnitude of spillovers}
Say we use Algorithm \ref{algo:sharp} and reject Hypothesis \ref{hyp:constant_direct_effect}. We may further wish to quantify the magnitude of these spillovers from treatment.  These methods can also be used to construct acceptance regions for more complex positive hypotheses about the size of spillovers in the network. To do this, we can use a similar test but with Equation \ref{eq:spillovers_frac} specifying a sharp null hypothesis given a choice of $\tau$ and $\rho$. We can, for example, use a test statistic that measures model fit (e.g., sum of squared residuals) \citep{bowers2016research} and determine a region of $\tau$ and $\rho$ values that we do not reject (i.e. an acceptance region). With only these two parameters, grid search is often feasible, but other search algorithms can be used. For more on this topic, see
\citet{bowers_reasoning_2013} and \citet{bowers2016research}.

The preceding methods require testing a sharp null hypothesis or a composite null consisting of a parametrically defined set of sharp nulls. In particular, we imposed the constant effects assumption that the direct effect of the treatment $\tau$ was common to all units. If direct effects are heterogeneous, these tests could reject the null even when there are no spillover effects of treatment. To partially address this concern, we could expand the null model to allow effects to be heterogeneous by observed subject covariates $X_i$; however, this would not allow for latent heterogeneity in direct effects. Outside the context of networks, we might be confident that, at least asymptotically, good choices of test statistics would result in tests that are not asymptotically sensitive to this heterogeneity \citep{chung2013exact}; however, we lack such asymptotic results for networks. In the next sections, we consider alternative methods that do not make use of these homogeneity assumptions. Nonetheless, the preceding methods may have some advantages in practice (e.g., greater power).

\subsubsection{Conditional randomization inference in networks}
How can we use randomization inference to test for spillovers without specifying the form of direct effects?
Consider a null hypothesis of no spillovers in the absence of assumptions about constant direct effects of treatment.
\begin{hypothesis}[No spillovers]
\label{hyp:no_spillovers}
$Y_i(z) = Y_i(z')$ for all $i \in V$, and all pairs of assignment vectors $z, z′ \in \{0,1\}^N$ such that $z_i$ = $z'_i$.
\end{hypothesis}
This hypothesis is not sharp because it does not specify how each subject would have behaved if its treatment were different. Rather, it posits levels sets of $Y_i(\cdot)$.
It is possible to test such non-sharp null hypotheses by using conditional randomization inference --- that is, by conditioning on functions of the treatment vector $Z$ \citep{rosenbaum1984conditional,aronow_general_2012, athey_exact_2017}.

Here consider the basic case of testing Hypothesis \ref{hyp:no_spillovers}. In particular, we can designate a subset of subjects as \emph{focal subjects} for which we examine their outcomes and condition on their observed treatment assignment \citep{aronow_general_2012, athey_exact_2017}. Note that, conditional on the focal subjects receiving the same treatment, Hypothesis \ref{hyp:no_spillovers} is now sharp for those subjects. We can implement this test as follows.

\begin{algo}[Conditional randomization inference for Hypothesis 2]
\label{algo:conditional}
Inputs: set of focal units $V_F \in V$, test statistic $T(\cdot, \cdot) : \mathbb{Y}^{|V_F|} \times \{0, 1\}^N$ that is a function of focal units’ outcomes and the treatment vector.
\begin{enumerate}
\item Draw permuted treatment vector $Z^*$ such that all focal units get the same treatment as observed, $Z^*_i = Z_i$ for all $i \in V_F$
\item Compute value of test statistic with observed outcomes and permuted treatment $T(Y_{V_F}, Z^*)$
\item Repeat 1 and 2 for R times, storing results as the $R$-vector $T_\text{null}$.
\item Compare observed and null test statistics, yielding
$$\text{p-value} = \frac{1}{R} \sum_{r = 1}^R \mathbbm{1} \{T(Y_{V_F}, Z) > T_{\text{null}, r} \}.$$
\end{enumerate}
\end{algo}
We would then reject Hypothesis \ref{hyp:no_spillovers}, and thus the stronger Hypothesis \ref{hyp:constant_direct_effect}, for small values of this p-value. This test has the correct Type I error rate without any assumptions about the model for direct effects.

How should the focal subjects be selected? Any choice is valid (i.e. results in correct Type I error rates), but this choice can affect power. First, in some cases, this choice may be obvious because of the availability of outcome data. For example, when joining treatment and network data with a second data set with outcomes, a researcher may only observe outcomes for a small fraction of subjects, which could then be designated the focal subjects \citep[e.g.,][]{jones2017social}. Second, theory or prior observations may suggest that some subject may not respond to social influence; it may be desirable to not include them as focal units.
Finally, the network itself can be used to select focal subjects to improve power \citep{athey_exact_2017,basse2016analyzing}.

It is possible to apply similar approaches to testing for higher-order spillovers, testing for spillovers on a second network, and other hypotheses about spillovers. When the null hypothesis allows for, e.g., spillovers from immediate neighbors on a relatively dense network, these methods may lack sufficient power to be useful. We refer readers to \citet{athey_exact_2017} for details. 

\subsubsection{Extension to edge-level treatments}
We have focused on the case where subjects, rather than edges, are assigned to treatments; however, similar methods can be used when edges are assigned as long as either (a) a sharp null hypothesis can be posited or (b) a non-sharp null hypothesis implies level sets that can be conditioned on.

\subsection{Estimating effects of peer behaviors}
Thus far we have described inference about effects of other subjects' randomly assigned treatments, while often the substantive questions are about effects of other subjects' behaviors (i.e. social influence). As noted above, if we assume that a subject's outcome is only affected by a peers' treatments via their behaviors, then evidence for spillovers from treatment is evidence for social influence. However, we are often interesting in quantifying the size of this social influence by, e.g., estimating effects of $D_j$ or $A_{ij}$ on $Y_i$.

We can proceed as before by considering the following sharp null hypothesis, which specifies how a subject's outcomes vary with its own treatment and peers' behaviors.

\begin{hypothesis}[Constant direct effects and social influence, ($\tau_0$, $\theta_0$)]
\label{hyp:constant_influence_sharp}
\begin{equation}
Y_i(z, d) = \tau_0 z_i + \theta_0 \sum_{j = 1}^N d_j \bar{A}_{ij} + \xi_i,
\end{equation}
for all $z \in \{0,1\}^N$, $d \in \dbm^N$, and $i \in v$.
\end{hypothesis}

According to Hypothesis \ref{hyp:constant_influence_sharp}, subjects are unaffected by others' treatments except as reflected in their neighbors behaviors $D_j$. This a \emph{complete mediation assumption} or \emph{exclusion restriction} and is encoded in Figure \ref{fig:diagram}. Combined with $Z$ having been randomized, this is sufficient for function of $Z$ to be used as instrumental variables for social influence. Following \citet{imbens2005robust}, we can then test Hypothesis \ref{hyp:constant_influence_sharp} by noting that it implies that $Y_i(z, d) - \tau_0 z_i - \theta_0 \sum_{j = 1}^N d_j \bar{A}_{ij}$ is invariant in $z$, and thus that Algorithm \ref{algo:sharp} can be applied with this alternative residualization of the outcomes.

\subsection{Other methods of analysis}
\label{sec:analysis_other}
There are some other methods available for statistical inference about spillovers and social influence with randomized experiments. Under monotonicity assumptions (i.e. that treating more subjects can only increase all subjects' potential outcomes) and with bounded outcomes, it is possible to construct confidence intervals for effects attributable to the observed treatment assignment \citep{choi_estimation_2014}. Or under local interference assumptions (i.e. subjects are only affected by immediate peers' treatments) and bounded degree, it is possible to do conservative asymptotic inference \citep{aronow2013estimating,laan_causal_2014}.

In some cases the presence of replication is helpful by allowing for plausible independence assumptions. First, there may be observation of multiple plausibly independent behaviors on a single network. For example, \cite{bakshy_social_2012} randomize a mechanism of social influence for many different subjects and brands. In their estimation and statistical inference, they assume that outcomes that do not have a common subject or brand are independent. They then use statistical methods that account for dependence of observations within brands and users \citep{bakshy2013uncertainty,cameron2011robust,owen2012bootstrapping}. Ignoring or not properly accounting for dependence in analyzing such experiments would increase the type I error rate.

Second, some networks consist of multiple sizable connected components (e.g., villages, schools), rather than a single giant component. However, often the lack of edges between components is an artifact of how the network is measured. For example, \citet{kim2015social} measure kinship and friendship relationship among rural villagers in Honduras, but edges between villages are not measured. On the other hand, \citet{cai_social_2015} measure inter-village edges, but nonetheless only allow for within-village dependence when conducting statistical inference. Thus, independence remains a potentially strong assumption.

\section{Interpretation and additional analyses} \label{sec:interpreting}

The simplest possible randomized experiment with a binary treatment (i.e. an ``A/B test'') could be used to estimate as little as a single causal parameter of interest --- the average treatment effect.  In many cases researchers have found that this is an unsatisfying conclusion to a study, especially given the costs of designing, planning, and implementing\footnote{One should realistically add to this list of costs, the expected cost of failure.  Researchers have not always succeeded in salvaging scientific knowledge from experiments and complexity of field experiments is associated with greater risk of unexpected problems.} randomized field experiments.
Therefore it is common for empirical researchers to conduct more extensive analysis of experimental data or to use it as input to models or simulations.  We have found that the results of field experiments, though exhibiting high internal and external validity, often motivate deeper questions about the underlying mechanism and alternative counterfactual questions that can be explored through modeling or simulation.

In Section~\ref{sec:analyzing}, we made assumptions about the structure of social influence and specified models or inferential procedures to detect or estimate it.  In most of these experiments we are more interested in how the effect scales with number or proportion of friends who engage in a particular behavior.  But beyond estimation of that response curve, there are two other broad types of questions that can be answered by experiments.  

The first is treatment effect heterogeneity --- the subpopulations of products, people, or social connections where social influence is stronger or weaker.  By fitting more complex models researchers can estimate heterogeneous treatment effects and these estimates can help suggest causal mechanisms or guide design of marketing efforts or public policies, analogous to finding predictors of positive response to clinical treatments in medicine.

The second is understanding optimal policies by simulating alternative policies.  Policy simulations can be used to extrapolate the results of randomized experiments to alternative policies which were never directly tested.  They are most commonly used by economists who have a rich history of using structural\footnote{Here we mean structural in the sense of imposing economic ``structure,'' meaning that assumptions about human behavior derived from theory are imposed in the models.} models in order to measure the effects of potential policy changes.

\subsection{Heterogeneous treatment effects}

One obvious type of effect heterogeneity is what clinical researchers might call a dose-response function, which characterizes how effects tend to scale as the number of friends who are influencing the person varies \citep{delean1978simultaneous}.  \cite{bakshy_social_2012} look at a slightly different dose-response function:  when an influential social cue from a single peer is present, how does the effect size vary with the tie-strength of that individual?  This heterogeneity is important to understand because it can inform advertising strategies.  For instance, knowing that close friends are far more influential than random friends, we might design a marketing campaign to encourage people to share with a small number of select friends rather than many of them.

Another common analysis is measuring how influence may be moderated by the demographic characteristics of the pair of people involved.  For instance, \cite{aral2012identifying} observe pairwise demographic attributes of the message sender and recipient and use this information to measure how the relative effectiveness of viral messages (Facebook notifications generated by app usage) varies as a means of identifying more influential or susceptible members of social networks.

It is completely plausible that the average treatment effect can be zero, yet obscure significant positive and negative treatment effects for many large subgroups that happen to cancel out.  \cite{taylor2014identity} show that the presence of some people's identity cues causes their content to receive higher and lower ratings than when their content is rendered anonymously. A distribution of effects that contains both positive and negative values is plausible in many social environments with fixed resources, such as status, reputation, or attention.

In all three of the aforementioned papers in this section, we would like to point out the effect heterogeneity does represent a ``free'' causal estimand.  If the experiment is designed to measure the ATE (the average treatment effect over the population), the effect heterogeneity we measure is simply an association between certain subgroups of the experiment and differential effects.  Researchers cannot make the claim that an intervention designed to move a subject from one subgroup to another would change their treatment effect.  For instance in \cite{taylor2014identity}, the experiment tests the effects of anonymization of a commenter's identity on ratings, but is unable to answer questions about what rating a person's content would receive if she had some alternate identity.  This type of counterfactual question is precisely the type of treatment effect heterogeneity that would drive policy decisions in the social advertising space. Such a finding can only be measured by a more complicated experiment which randomizes \emph{which} identity is presented among some set of choices -- a much more difficult experiment to design and implement.

We end this section with a note of warning about seeking results based on treatment effect heterogeneity.  As researchers search dimensions by which treatment effects may exhibit differential effects, they may increase the rate of false discovery.  Independently testing many heterogeneity on many possible dimensions, or for many subgroups of the experiment will invariably result in false positive results as one of the subgroups may be ``lucky.''\footnote{See for example: \url{https://xkcd.com/882/}}  There are reasonable methods to control this risk while still detecting interesting heterogeneity, see \cite{athey2015machine} for a detailed discussion and recent methodological development.

\subsection{Policy simulations}

Given the obvious importance of experiments for effective policy decisions, it is natural to ask for a policy recommendation at the conclusion of a study based on a randomized experiment.  Often a policy recommendation is not directly recoverable from causal quantities of interest.  For instance, the average treatment effect (ATE) might tell you that the treatment has a positive effect on some outcome \emph{on average}, but it does not necessarily follow that everyone should receive the treatment.  Treatments have costs which might need to be weighed and nature of social spillovers means that treating a friend of an individual can be a substitute or a complement for treating that individual directly.

\cite{ryan2012heterogeneity} report policy simulations, employing models containing economic structure based on assumptions about individual behavior in the presence of peer effects \citep{hartmann2010demand}.  The key idea behind the policy simulation approach is that the experiment is used to estimate parameters of the model and then the model can be used to extrapolate the findings to more complex or interesting policies than those randomly set in the original data set.   Policy simulations are often used in conjunction with natural experiment, where the researcher did not ex ante decide the most informative randomization and would like to answer some additional questions at the cost of imposing additional assumptions through a model.

Another example of reporting policy simulations is \cite{aral2013engineering}, which applies experimental estimates from \cite{aral2012identifying} to form the basis of an optimal seeding strategy in networks.  As a key feature, their experiment estimated the degree to which influence and susceptibility to influence were clustered in the network, which is an important feature for understanding diffusion processes.

\section{Conclusion} \label{sec:conclusion}

We believe that valid causal inference is an important goal in the practice of social science.  There is obvious utility in knowing causal structure --- good policy decisions \emph{require} that the policy-maker at least know the sign of a causal effect.  But also from a purely scientific perspective, measurements which do not have causal interpretation lack usefulness and insight because they afford multiple explanations. Correlations are interesting, but they usually cannot uniquely identify an explanation for a social phenomenon.

We admit there is perhaps a bit of experimental dogma present in the social sciences and it is often possible to satisfyingly answer questions through some combination of reasonable assumptions, models, and observational data.  However, the realm of social influence is one where alternative explanations are difficult to rule out without some exogenous variation which can identify causal effects. \cite{manzi2012uncontrolled} refers to this problematic aspect of human behavior as ``high causal density.''  In domains of high causal density, where there are highly dense causal graphs that can explain the observed associations in data we collect, credible causal claims require randomization, either by the researcher or by nature.

%It has been the goal of this review to provide a simple framework for thinking about randomized experiments in the domain of social influence.  The framework we have outlined applies to dozens of randomized experiments and is extensible to many other possible studies.  We envision that with this framework and references in hand, researchers will be well equipped to design experiments that can help detect and estimate social influence across a variety of fields and domains.

\begin{acknowledgement}
We would like to thank Lada Adamic, Norberto Andrade, Eytan Bakshy, and George Berry for helpful discussions in preparing this review.
\end{acknowledgement}

\bibliographystyle{spbasic}
\bibliography{references}

\begin{thebibliography}{121}
\providecommand{\natexlab}[1]{#1}
\providecommand{\url}[1]{{#1}}
\providecommand{\urlprefix}{URL }
\expandafter\ifx\csname urlstyle\endcsname\relax
  \providecommand{\doi}[1]{DOI~\discretionary{}{}{}#1}\else
  \providecommand{\doi}{DOI~\discretionary{}{}{}\begingroup
  \urlstyle{rm}\Url}\fi
\providecommand{\eprint}[2][]{\url{#2}}

\bibitem[{Anagnostopoulos et~al(2008)Anagnostopoulos, Kumar, and
  Mahdian}]{anagnostopoulos2008influence}
Anagnostopoulos A, Kumar R, Mahdian M (2008) Influence and correlation in
  social networks. In: Proceedings of the 14th ACM SIGKDD international
  conference on Knowledge discovery and data mining, ACM, pp 7--15

\bibitem[{Angelucci and De~Giorgi(2009)}]{angelucci_indirect_2009}
Angelucci M, De~Giorgi G (2009) Indirect effects of an aid program: How do cash
  transfers affect ineligibles' consumption? American Economic Review
  99(1):486--508

\bibitem[{Angrist(2014)}]{angrist_perils_2014}
Angrist JD (2014) The perils of peer effects. Labour Economics 30:98--108

\bibitem[{Aplin et~al(2015)Aplin, Farine, Morand-Ferron, Cockburn, Thornton,
  and Sheldon}]{aplin_experimentally_2015}
Aplin LM, Farine DR, Morand-Ferron J, Cockburn A, Thornton A, Sheldon BC (2015)
  Experimentally induced innovations lead to persistent culture via conformity
  in wild birds. Nature 518(7540):538--541

\bibitem[{Aral(2016)}]{aral_networked_2016}
Aral S (2016) Networked experiments. In: The Oxford Handbook of the Economics
  of Networks, Oxford University Press

\bibitem[{Aral and Walker(2011)}]{aral2011creating}
Aral S, Walker D (2011) Creating social contagion through viral product design:
  A randomized trial of peer influence in networks. Management science
  57(9):1623--1639

\bibitem[{Aral and Walker(2012)}]{aral2012identifying}
Aral S, Walker D (2012) Identifying influential and susceptible members of
  social networks. Science 337(6092):337--341

\bibitem[{Aral and Walker(2014)}]{aral2014tie}
Aral S, Walker D (2014) Tie strength, embeddedness, and social influence: A
  large-scale networked experiment. Management Science 60(6):1352--1370

\bibitem[{Aral et~al(2009)Aral, Muchnik, and
  Sundararajan}]{aral2009distinguishing}
Aral S, Muchnik L, Sundararajan A (2009) Distinguishing influence-based
  contagion from homophily-driven diffusion in dynamic networks. Proceedings of
  the National Academy of Sciences 106(51):21,544--21,549

\bibitem[{Aral et~al(2013)Aral, Muchnik, and
  Sundararajan}]{aral2013engineering}
Aral S, Muchnik L, Sundararajan A (2013) Engineering social contagions: Optimal
  network seeding in the presence of homophily. Network Science 1(02):125--153

\bibitem[{Aronow(2012)}]{aronow_general_2012}
Aronow PM (2012) A general method for detecting interference between units in
  randomized experiments. Sociological Methods \& Research 41(1):3--16

\bibitem[{Aronow and Samii(2017)}]{aronow2013estimating}
Aronow PM, Samii C (2017) Estimating average causal effects under general
  interference, with application to a social network experiment. Annals of
  Applied Statistics Forthcoming

\bibitem[{Asch(1955)}]{asch1955opinions}
Asch SE (1955) Opinions and social pressure. Readings about the social animal
  193:17--26

\bibitem[{Athey and Imbens(2015)}]{athey2015machine}
Athey S, Imbens GW (2015) Machine learning methods for estimating heterogeneous
  causal effects. stat 1050(5)

\bibitem[{Athey and Imbens(2017)}]{athey2017econometrics}
Athey S, Imbens GW (2017) The econometrics of randomized experiments. Handbook
  of Economic Field Experiments 1:73--140

\bibitem[{Athey et~al(2017)Athey, Eckles, and Imbens}]{athey_exact_2017}
Athey S, Eckles D, Imbens GW (2017) Exact p-values for network interference.
  Journal of the American Statistical Association Forthcoming.

\bibitem[{Backstrom and Leskovec(2011)}]{backstrom2011supervised}
Backstrom L, Leskovec J (2011) Supervised random walks: Predicting and
  recommending links in social networks. In: Proceedings of the fourth ACM
  international conference on Web search and data mining, ACM, pp 635--644

\bibitem[{Baird et~al(2016)Baird, Bohren, McIntosh, and
  Ozler}]{baird2016optimal}
Baird S, Bohren JA, McIntosh C, Ozler B (2016) Optimal design of experiments in
  the presence of interference,
  \urlprefix\url{https://papers.ssrn.com/sol3/papers.cfm?abstract_id=2900967},
  working paper

\bibitem[{Bakshy and Eckles(2013)}]{bakshy2013uncertainty}
Bakshy E, Eckles D (2013) Uncertainty in online experiments with dependent
  data: an evaluation of bootstrap methods. In: Proceedings of the 19th ACM
  SIGKDD international conference on Knowledge discovery and data mining, ACM,
  pp 1303--1311

\bibitem[{Bakshy et~al(2012{\natexlab{a}})Bakshy, Eckles, Yan, and
  Rosenn}]{bakshy_social_2012}
Bakshy E, Eckles D, Yan R, Rosenn I (2012{\natexlab{a}}) Social influence in
  social advertising: Evidence from field experiments. In: Proceedings of the
  {ACM} conference on Electronic Commerce, {ACM}

\bibitem[{Bakshy et~al(2012{\natexlab{b}})Bakshy, Rosenn, Marlow, and
  Adamic}]{bakshy_role_2011}
Bakshy E, Rosenn I, Marlow C, Adamic L (2012{\natexlab{b}}) The role of social
  networks in information diffusion. In: Proceedings of the 21st international
  conference on World Wide Web, ACM, WWW '12, pp 519--528

\bibitem[{Bakshy et~al(2014)Bakshy, Eckles, and
  Bernstein}]{bakshy_designing_2014}
Bakshy E, Eckles D, Bernstein MS (2014) Designing and deploying online field
  experiments. In: Proceedings of the 23rd international conference on World
  Wide Web, pp 283--292

\bibitem[{Bakshy et~al(2015)Bakshy, Messing, and Adamic}]{bakshy2015exposure}
Bakshy E, Messing S, Adamic LA (2015) Exposure to ideologically diverse news
  and opinion on facebook. Science 348(6239):1130--1132

\bibitem[{Bapna and Umyarov(2015)}]{bapna2015your}
Bapna R, Umyarov A (2015) Do your online friends make you pay? {A} randomized
  field experiment on peer influence in online social networks. Management
  Science 61(8):1902--1920

\bibitem[{Bass(1969)}]{bass1969ms}
Bass FM (1969) A new product growth for model consumer durables. Management
  Science 15(5):215--227, \doi{10.2307/2628128}

\bibitem[{Basse and Feller(2017)}]{basse2016analyzing}
Basse G, Feller A (2017) Analyzing multilevel experiments in the presence of
  peer effects. Journal of the American Statistical Association Forthcoming

\bibitem[{Basse and Airoldi(2015)}]{basse2015optimal}
Basse GW, Airoldi EM (2015) Optimal model-assisted design of experiments for
  network correlated outcomes suggests new notions of network balance. arXiv
  preprint arXiv:150700803

\bibitem[{Beaman et~al(2015)Beaman, BenYishay, Magruder, and
  Mobarak}]{beaman_can_2015}
Beaman L, BenYishay A, Magruder J, Mobarak AM (2015) Can network theory-based
  targeting increase technology adoption?, working paper

\bibitem[{Beasley and Mason(2015)}]{beasley2015emotional}
Beasley A, Mason W (2015) Emotional states vs. emotional words in social media.
  In: Proceedings of the ACM Web Science Conference, ACM, p~31

\bibitem[{Berry and Taylor(2017)}]{berry2017discussion}
Berry G, Taylor SJ (2017) Discussion quality diffuses in the digital public
  square. In: Proceedings of the 26th International Conference on World Wide
  Web, International World Wide Web Conferences Steering Committee, pp
  1371--1380

\bibitem[{Bond et~al(2012)Bond, Fariss, Jones, Kramer, Marlow, Settle, and
  Fowler}]{bond_experiment_2012}
Bond RM, Fariss CJ, Jones JJ, Kramer AD, Marlow C, Settle JE, Fowler JH (2012)
  A 61-million-person experiment in social influence and political
  mobilization. Nature 489(7415):295--298

\bibitem[{Bond et~al(2016)Bond, Settle, Fariss, Jones, and
  Fowler}]{bond2016social}
Bond RM, Settle JE, Fariss CJ, Jones JJ, Fowler JH (2016) Social endorsement
  cues and political participation. Political Communication pp 1--21

\bibitem[{Bowers et~al(2013)Bowers, Fredrickson, and
  Panagopoulos}]{bowers_reasoning_2013}
Bowers J, Fredrickson MM, Panagopoulos C (2013) Reasoning about interference
  between units: A general framework. Political Analysis 21(1):97--124

\bibitem[{Bowers et~al(2016)Bowers, Fredrickson, and
  Aronow}]{bowers2016research}
Bowers J, Fredrickson MM, Aronow PM (2016) A more powerful test statistic for
  reasoning about interference between units. Political Analysis p mpw018

\bibitem[{Cai et~al(2015)Cai, De~Janvry, and Sadoulet}]{cai_social_2015}
Cai J, De~Janvry A, Sadoulet E (2015) Social networks and the decision to
  insure. American Economic Journal: Applied Economics 7(2):81--108

\bibitem[{Cameron et~al(2011)Cameron, Gelbach, and Miller}]{cameron2011robust}
Cameron AC, Gelbach JB, Miller DL (2011) Robust inference with multiway
  clustering. Journal of Business \& Economic Statistics 29(2):238--249

\bibitem[{Carrell et~al(2009)Carrell, Fullerton, and West}]{carrell_does_2009}
Carrell SE, Fullerton RL, West JE (2009) Does your cohort matter? {M}easuring
  peer effects in college achievement. Journal of Labor Economics
  27(3):439--464

\bibitem[{Carrell et~al(2011)Carrell, Sacerdote, and
  West}]{carrell_natural_2011}
Carrell SE, Sacerdote BI, West JE (2011) From natural variation to optimal
  policy? {T}he lucas critique meets peer effects. National Bureau of Economic
  Research Working Paper Series No. 16865,
  \urlprefix\url{http://www.nber.org/papers/w16865}

\bibitem[{Centola(2010)}]{centola2010spread}
Centola D (2010) The spread of behavior in an online social network experiment.
  science 329(5996):1194--1197

\bibitem[{Choi(2017)}]{choi_estimation_2014}
Choi DS (2017) Estimation of monotone treatment effects in network experiments.
  Journal of the American Statistical Association Forthcoming

\bibitem[{Chung and Romano(2013)}]{chung2013exact}
Chung E, Romano JP (2013) Exact and asymptotically robust permutation tests.
  The Annals of Statistics 41(2):484--507

\bibitem[{Coey and Bailey(2016)}]{coey2016people}
Coey D, Bailey M (2016) People and cookies: Imperfect treatment assignment in
  online experiments. In: Proceedings of the 25th International Conference on
  World Wide Web, International World Wide Web Conferences Steering Committee,
  pp 1103--1111

\bibitem[{Coppock et~al(2015)Coppock, Guess, and
  Ternovski}]{coppock_treatments_2015}
Coppock A, Guess A, Ternovski J (2015) When treatments are tweets: A network
  mobilization experiment over twitter. Political Behavior pp 1--24

\bibitem[{Currarini et~al(2010)Currarini, Jackson, and
  Pin}]{currarini2010identifying}
Currarini S, Jackson MO, Pin P (2010) Identifying the roles of race-based
  choice and chance in high school friendship network formation. Proceedings of
  the National Academy of Sciences 107(11):4857--4861

\bibitem[{De~Choudhury et~al(2010)De~Choudhury, Mason, Hofman, and
  Watts}]{de2010inferring}
De~Choudhury M, Mason WA, Hofman JM, Watts DJ (2010) Inferring relevant social
  networks from interpersonal communication. In: Proceedings of the 19th
  international conference on World wide web, ACM, pp 301--310

\bibitem[{DeLean et~al(1978)DeLean, Munson, and
  Rodbard}]{delean1978simultaneous}
DeLean A, Munson P, Rodbard D (1978) Simultaneous analysis of families of
  sigmoidal curves: application to bioassay, radioligand assay, and
  physiological dose-response curves. American Journal of
  Physiology-Gastrointestinal and Liver Physiology 235(2):G97--102

\bibitem[{Dittrich et~al(2012)Dittrich, Kenneally et~al}]{dittrich2012menlo}
Dittrich D, Kenneally E, et~al (2012) The {M}enlo {R}eport: Ethical principles
  guiding information and communication technology research. US Department of
  Homeland Security

\bibitem[{Duflo and Saez(2003)}]{duflo_role_2003}
Duflo E, Saez E (2003) The role of information and social interactions in
  retirement plan decisions: Evidence from a randomized experiment. Quarterly
  Journal of Economics 118(3):815--842

\bibitem[{Eckles et~al(2016)Eckles, Kizilcec, and
  Bakshy}]{eckles_estimating_2016}
Eckles D, Kizilcec RF, Bakshy E (2016) Estimating peer effects in networks with
  peer encouragement designs. Proceedings of the National Academy of Sciences
  113(27):7316--7322

\bibitem[{Eckles et~al(2017)Eckles, Karrer, and Ugander}]{eckles_design_2017}
Eckles D, Karrer B, Ugander J (2017) Design and analysis of experiments in
  networks: Reducing bias from interference. Journal of Causal Inference

\bibitem[{Firth et~al(2016)Firth, Sheldon, and Farine}]{firth_pathways_2016}
Firth JA, Sheldon BC, Farine DR (2016) Pathways of information transmission
  among wild songbirds follow experimentally imposed changes in social foraging
  structure. Biology Letters 12(6):20160,144

\bibitem[{Fisher(1935)}]{fisher1935design}
Fisher RA (1935) The Design of Experiments. Oliver and Boyd, Edinburgh

\bibitem[{Forastiere et~al(2015)Forastiere, Mealli, and
  VanderWeele}]{forastiere_identification_2015}
Forastiere L, Mealli F, VanderWeele TJ (2015) Identification and estimation of
  causal mechanisms in clustered encouragement designs: Disentangling bed nets
  using {B}ayesian principal stratification. Journal of the American
  Statistical Association (just-accepted):1--44

\bibitem[{Gerber and Green(2012)}]{gerber2012field}
Gerber AS, Green DP (2012) Field Experiments: Design, Analysis, and
  Interpretation. WW Norton

\bibitem[{Granovetter(1978)}]{granovetter1978threshold}
Granovetter M (1978) {Threshold models of collective behavior}. The American
  Journal of Sociology 83(6):1420--1443

\bibitem[{Halberstam and Knight(2016)}]{halberstam2016homophily}
Halberstam Y, Knight B (2016) Homophily, group size, and the diffusion of
  political information in social networks: Evidence from twitter. Journal of
  Public Economics 143:73--88

\bibitem[{Halloran and Hudgens(2016)}]{halloran2016dependent}
Halloran ME, Hudgens MG (2016) Dependent happenings: A recent methodological
  review. Current epidemiology reports 3(4):297--305

\bibitem[{Hartmann(2010)}]{hartmann2010demand}
Hartmann WR (2010) Demand estimation with social interactions and the
  implications for targeted marketing. Marketing Science 29(4):585--601

\bibitem[{Hasan and Koning(2017)}]{hasan2017conversational}
Hasan S, Koning R (2017) Conversational peers and idea generation: Evidence
  from a field experiment. under Review

\bibitem[{Higgins et~al(2016)Higgins, S{\"a}vje, and
  Sekhon}]{higgins2016improving}
Higgins MJ, S{\"a}vje F, Sekhon JS (2016) Improving massive experiments with
  threshold blocking. Proceedings of the National Academy of Sciences
  113(27):7369--7376

\bibitem[{Huber and Malhotra(2017)}]{huber2017political}
Huber GA, Malhotra N (2017) Political homophily in social relationships:
  Evidence from online dating behavior. The Journal of Politics 79(1):269--283

\bibitem[{Imbens and Rosenbaum(2005)}]{imbens2005robust}
Imbens GW, Rosenbaum PR (2005) Robust, accurate confidence intervals with a
  weak instrument: Quarter of birth and education. Journal of the Royal
  Statistical Society: Series A (Statistics in Society) 168(1):109--126

\bibitem[{Imbens and Rubin(2015)}]{imbens2015causal}
Imbens GW, Rubin DB (2015) Causal Inference in Statistics, Social, and
  Biomedical Sciences. Cambridge University Press

\bibitem[{Jackman and Kanerva(2016)}]{jackman2016evolving}
Jackman M, Kanerva L (2016) Evolving the irb: Building robust review for
  industry research. Washington and Lee Law Review Online 72(3):442

\bibitem[{Jones et~al(2013)Jones, Settle, Bond, Fariss, Marlow, and
  Fowler}]{jones2013inferring}
Jones JJ, Settle JE, Bond RM, Fariss CJ, Marlow C, Fowler JH (2013) Inferring
  tie strength from online directed behavior. PloS one 8(1):e52,168

\bibitem[{Jones et~al(2017)Jones, Bond, Bakshy, Eckles, and
  Fowler}]{jones2017social}
Jones JJ, Bond RM, Bakshy E, Eckles D, Fowler JH (2017) Social influence and
  political mobilization: Further evidence from a randomized experiment in the
  2012 {US} presidential election. PLoS ONE 12(4):e0173,851

\bibitem[{Karrer and Newman(2011)}]{karrer_stochastic_2011}
Karrer B, Newman ME (2011) Stochastic blockmodels and community structure in
  networks. Physical Review E 83(1):016,107

\bibitem[{Kearns et~al(2006)Kearns, Suri, and
  Montfort}]{kearns2006experimental}
Kearns M, Suri S, Montfort N (2006) An experimental study of the coloring
  problem on human subject networks. Science 313(5788):824--827

\bibitem[{Kempe et~al(2003)Kempe, Kleinberg, and Tardos}]{kempe2003maximizing}
Kempe D, Kleinberg J, Tardos {\'E} (2003) Maximizing the spread of influence
  through a social network. In: Proceedings of the ninth ACM SIGKDD
  international conference on Knowledge discovery and data mining, ACM, pp
  137--146

\bibitem[{Kim et~al(2015)Kim, Hwong, Stafford, Hughes, O'Malley, Fowler, and
  Christakis}]{kim2015social}
Kim DA, Hwong AR, Stafford D, Hughes DA, O'Malley AJ, Fowler JH, Christakis NA
  (2015) Social network targeting to maximise population behaviour change: a
  cluster randomised controlled trial. The Lancet 386(9989):145--153

\bibitem[{Kohavi et~al(2012)Kohavi, Deng, Frasca, Longbotham, Walker, and
  Xu}]{kohavi2012trustworthy}
Kohavi R, Deng A, Frasca B, Longbotham R, Walker T, Xu Y (2012) Trustworthy
  online controlled experiments: Five puzzling outcomes explained. In:
  Proceedings of the 18th ACM SIGKDD international conference on Knowledge
  discovery and data mining, ACM, pp 786--794

\bibitem[{Kossinets and Watts(2006)}]{kossinets2006empirical}
Kossinets G, Watts DJ (2006) Empirical analysis of an evolving social network.
  Science 311(5757):88--90

\bibitem[{Kramer et~al(2014)Kramer, Guillory, and
  Hancock}]{kramer2014experimental}
Kramer AD, Guillory JE, Hancock JT (2014) Experimental evidence of
  massive-scale emotional contagion through social networks. Proceedings of the
  National Academy of Sciences 111(24):8788--8790

\bibitem[{van~der Laan(2014)}]{laan_causal_2014}
van~der Laan MJ (2014) Causal inference for a population of causally connected
  units. Journal of Causal Inference pp 1--62

\bibitem[{Leider et~al(2009)Leider, M{\"o}bius, Rosenblat, and
  Do}]{leider2009directed}
Leider S, M{\"o}bius MM, Rosenblat T, Do QA (2009) Directed altruism and
  enforced reciprocity in social networks. The Quarterly Journal of Economics
  124(4):1815--1851

\bibitem[{Leung(2017)}]{leung_weak_2017}
Leung MP (2017) A weak law for moments of pairwise-stable networks, working
  paper. http://dx.doi.org/10.2139/ssrn.2663685

\bibitem[{Manski(1993)}]{manski1993identification}
Manski CF (1993) Identification of endogenous social effects: The reflection
  problem. The review of economic studies 60(3):531--542

\bibitem[{Manski(2000)}]{manski2000economic}
Manski CF (2000) Economic analysis of social interactions. Tech. rep., National
  bureau of economic research

\bibitem[{Manzi(2012)}]{manzi2012uncontrolled}
Manzi J (2012) Uncontrolled: The surprising payoff of trial-and-error for
  business, politics, and society. Basic Books

\bibitem[{Marmaros and Sacerdote(2006)}]{marmaros2006friendships}
Marmaros D, Sacerdote B (2006) How do friendships form? The Quarterly Journal
  of Economics 121(1):79--119

\bibitem[{Mason and Suri(2012)}]{mason2012conducting}
Mason W, Suri S (2012) Conducting behavioral research on amazon’s mechanical
  turk. Behavior research methods 44(1):1--23

\bibitem[{Mason and Watts(2012)}]{mason2012collaborative}
Mason W, Watts DJ (2012) Collaborative learning in networks. Proceedings of the
  National Academy of Sciences 109(3):764--769

\bibitem[{McPherson et~al(2001)McPherson, Smith-Lovin, and
  Cook}]{mcpherson2001birds}
McPherson M, Smith-Lovin L, Cook JM (2001) Birds of a feather: Homophily in
  social networks. Annual review of sociology 27(1):415--444

\bibitem[{Miguel and Kremer(2004)}]{miguel_worms:_2004}
Miguel E, Kremer M (2004) Worms: Identifying impacts on education and health in
  the presence of treatment externalities. Econometrica 72(1):159--217

\bibitem[{Miratrix et~al(2013)Miratrix, Sekhon, and Yu}]{miratrix2013adjusting}
Miratrix LW, Sekhon JS, Yu B (2013) Adjusting treatment effect estimates by
  post-stratification in randomized experiments. Journal of the Royal
  Statistical Society: Series B (Statistical Methodology) 75(2):369--396

\bibitem[{Moffitt et~al(2001)}]{moffitt2001policy}
Moffitt RA, et~al (2001) Policy interventions, low-level equilibria, and social
  interactions. Social dynamics 4:45--82

\bibitem[{Morris(2000)}]{morris2000contagion}
Morris S (2000) {Contagion}. The Review of Economic Studies 67(1):57

\bibitem[{Muchnik et~al(2013)Muchnik, Aral, and Taylor}]{muchnik2013social}
Muchnik L, Aral S, Taylor SJ (2013) Social influence bias: A randomized
  experiment. Science 341(6146):647--651

\bibitem[{Narayanan and Shmatikov(2010)}]{narayanan2010myths}
Narayanan A, Shmatikov V (2010) Myths and fallacies of personally identifiable
  information. Communications of the ACM 53(6):24--26

\bibitem[{Nickerson(2008)}]{nickerson2008voting}
Nickerson DW (2008) Is voting contagious? {E}vidence from two field
  experiments. American Political Science Review 102(01):49--57

\bibitem[{Ohm(2010)}]{ohm2010broken}
Ohm P (2010) Broken promises of privacy: Responding to the surprising failure
  of anonymization. UCLA L Rev 57:1701--1819

\bibitem[{Owen and Eckles(2012)}]{owen2012bootstrapping}
Owen AB, Eckles D (2012) Bootstrapping data arrays of arbitrary order. The
  Annals of Applied Statistics 6(3):895--927

\bibitem[{Paluck et~al(2016)Paluck, Shepherd, and
  Aronow}]{paluck_changing_2016}
Paluck EL, Shepherd H, Aronow PM (2016) Changing climates of conflict: A social
  network experiment in 56 schools. Proceedings of the National Academy of
  Sciences 113(3):566--571

\bibitem[{Rand et~al(2011)Rand, Arbesman, and Christakis}]{rand2011dynamic}
Rand DG, Arbesman S, Christakis NA (2011) Dynamic social networks promote
  cooperation in experiments with humans. Proceedings of the National Academy
  of Sciences 108(48):19,193--19,198

\bibitem[{Rand et~al(2014)Rand, Nowak, Fowler, and Christakis}]{rand2014static}
Rand DG, Nowak MA, Fowler JH, Christakis NA (2014) Static network structure can
  stabilize human cooperation. Proceedings of the National Academy of Sciences
  111(48):17,093--17,098

\bibitem[{Resea and Ryan(1978)}]{national1978belmont}
Resea B, Ryan KJP (1978) The Belmont Report: Ethical Principles and Guidelines
  for the Protection of Human Subjects of Research-the National Commission for
  the Protection of Human Subjects of Biomedical and Behavioral Research. US
  Government Printing Office

\bibitem[{Resnick et~al(1997)Resnick, Bearman, Blum, Bauman, Harris, Jones,
  Tabor, Beuhring, Sieving, Shew et~al}]{resnick1997protecting}
Resnick MD, Bearman PS, Blum RW, Bauman KE, Harris KM, Jones J, Tabor J,
  Beuhring T, Sieving RE, Shew M, et~al (1997) Protecting adolescents from
  harm: Findings from the {N}ational {L}ongitudinal {S}tudy on {A}dolescent
  {H}ealth. JAMA 278(10):823--832

\bibitem[{Riach and Rich(2004)}]{riach2004deceptive}
Riach PA, Rich J (2004) Deceptive field experiments of discrimination: are they
  ethical? Kyklos 57(3):457--470

\bibitem[{Rock et~al(2016)Rock, Aral, and Taylor}]{rock2016identification}
Rock D, Aral S, Taylor SJ (2016) Identification of peer effects in networked
  panel data. In: Thirty Seventh International Conference on Information
  Systems

\bibitem[{Rosenbaum(1984)}]{rosenbaum1984conditional}
Rosenbaum PR (1984) Conditional permutation tests and the propensity score in
  observational studies. Journal of the American Statistical Association
  79(387):565--574

\bibitem[{Ryan and Gross(1943)}]{ryan1943diffusion}
Ryan B, Gross NC (1943) The diffusion of hybrid seed corn in two {I}owa
  communities. Rural Sociology 8(1):15

\bibitem[{Ryan and Tucker(2012)}]{ryan2012heterogeneity}
Ryan SP, Tucker C (2012) Heterogeneity and the dynamics of technology adoption.
  Quantitative Marketing and Economics 10(1):63--109

\bibitem[{Sacerdote(2001)}]{sacerdote2001peer}
Sacerdote B (2001) Peer effects with random assignment: Results for dartmouth
  roommates. The Quarterly journal of economics 116(2):681--704

\bibitem[{Salganik(2017)}]{salganik2017bit}
Salganik MJ (2017) Bit by Bit: Social Research in the Digital Age. Princeton
  University Press

\bibitem[{Saul et~al(2017)Saul, Hudgens, and Halloran}]{saul2017causal}
Saul BC, Hudgens MG, Halloran ME (2017) Causal inference in the study of
  infectious disease. Handbook of Statistics

\bibitem[{Schelling(1969)}]{schelling1969models}
Schelling T (1969) {Models of segregation}. The American Economic Review
  59(2):488--493

\bibitem[{Schelling(1971)}]{schelling1971dynamic}
Schelling T (1971) {Dynamic models of segregation}. The Journal of Mathematical
  Sociology 1(2):143--186

\bibitem[{Schelling(1973)}]{schelling1973hockey}
Schelling T (1973) {Hockey helmets, concealed weapons, and daylight saving: A
  study of binary choices with externalities}. The Journal of Conflict
  Resolution 17(3):381--428

\bibitem[{Schultz et~al(2012)Schultz, Piepgrass, Weng, Ferrante, Verma,
  Martinazzi, Alison, and Mao}]{schultz2012methods}
Schultz AP, Piepgrass B, Weng CC, Ferrante D, Verma D, Martinazzi P, Alison T,
  Mao Z (2012) Methods and systems for determining use and content of {PYMK}
  based on value model. US Patent App. 13/659,695

\bibitem[{Shalita et~al(2016)Shalita, Karrer, Kabiljo, Sharma, Presta, Adcock,
  Kllapi, and Stumm}]{shalita2016social}
Shalita A, Karrer B, Kabiljo I, Sharma A, Presta A, Adcock A, Kllapi H, Stumm M
  (2016) Social {H}ash: An assignment framework for optimizing distributed
  systems operations on social networks. In: NSDI, pp 455--468

\bibitem[{Shalizi and Thomas(2011)}]{shalizi2011homophily}
Shalizi CR, Thomas AC (2011) Homophily and contagion are generically confounded
  in observational social network studies. Sociological methods \& research
  40(2):211--239

\bibitem[{Suri and Watts(2011)}]{suri2011cooperation}
Suri S, Watts DJ (2011) Cooperation and contagion in web-based, networked
  public goods experiments. PloS one 6(3):e16,836

\bibitem[{Taylor et~al(2013)Taylor, Bakshy, and Aral}]{taylor2013selection}
Taylor SJ, Bakshy E, Aral S (2013) Selection effects in online sharing:
  Consequences for peer adoption. In: Proceedings of the fourteenth ACM
  conference on Electronic commerce, ACM, pp 821--836

\bibitem[{Taylor et~al(2014)Taylor, Muchnik, and Aral}]{taylor2014identity}
Taylor SJ, Muchnik L, Aral S (2014) Identity and opinion: A randomized
  experiment, working paper. http://dx.doi.org/10.2139/ssrn.2538130

\bibitem[{Tchetgen et~al(2017)Tchetgen, Fulcher, and
  Shpitser}]{tchetgen2017auto}
Tchetgen EJT, Fulcher I, Shpitser I (2017) Auto-{G}-computation of causal
  effects on a network. arXiv preprint arXiv:170901577

\bibitem[{Toulis and Kao(2013)}]{toulis_estimation_2013}
Toulis P, Kao E (2013) Estimation of causal peer influence effects. In:
  Proceedings of The 30th International Conference on Machine Learning, pp
  1489--1497

\bibitem[{Ugander et~al(2013)Ugander, Karrer, Backstrom, and
  Kleinberg}]{ugander_graph_2013}
Ugander J, Karrer B, Backstrom L, Kleinberg JM (2013) Graph cluster
  randomization: {N}etwork exposure to multiple universes. In: Proc. of KDD,
  {ACM}

\bibitem[{Valente(1996)}]{valente1996social}
Valente TW (1996) Social network thresholds in the diffusion of innovations.
  Social networks 18(1):69--89

\bibitem[{VanderWeele et~al(2012)VanderWeele, Tchetgen, and
  Halloran}]{vanderweele2012components}
VanderWeele TJ, Tchetgen EJT, Halloran ME (2012) Components of the indirect
  effect in vaccine trials: Identification of contagion and infectiousness
  effects. Epidemiology (Cambridge, Mass) 23(5):751

\bibitem[{Walker and Muchnik(2014)}]{walker_design_2014}
Walker D, Muchnik L (2014) Design of randomized experiments in networks.
  Proceedings of the IEEE 102(12):1940--1951

\bibitem[{Xu et~al(2015)Xu, Chen, Fernandez, Sinno, and
  Bhasin}]{xu2015infrastructure}
Xu Y, Chen N, Fernandez A, Sinno O, Bhasin A (2015) From infrastructure to
  culture: {A/B} testing challenges in large scale social networks. In:
  Proceedings of the 21th ACM SIGKDD International Conference on Knowledge
  Discovery and Data Mining, ACM, pp 2227--2236

\end{thebibliography}

\end{document}